\documentclass[11pt]{article}

\usepackage{amsthm}
\usepackage{amsmath,amssymb}
\usepackage{xspace}
\usepackage{color}
\usepackage{epsfig}
\graphicspath{{.}{./figures/}}

\usepackage{macros/tikzfig}
\usepackage{keycommand}
\usepackage{hyperref}

\newkeycommand{\boxmap}[style=small box][1]{\,\tikz{\node[style=\commandkey{style}] (x) {$#1$};\draw(0,-1.25)--(x)--(0,1.25);}\,}
\newkeycommand{\dboxmap}[style=dbox][1]{\,\tikz{\node[style=\commandkey{style}] (x) {$#1$};\draw[boldedge] (0,-1.25)--(x)--(0,1.25);}\,}

\newkeycommand{\wirelabel}[style=white label]{\,\tikz{\node[style=\commandkey{style}]{};}\,\xspace}

\newkeycommand{\pointmap}[style=point][1]{\,\tikz{\node[style=\commandkey{style}] (x) at (0,-0.3) {$#1$};\node [style=none] (3) at (0, 0.9) {};\node [style=none] (3) at (0, -0.9) {};\draw (x)--(0,0.7);}\,}

\newkeycommand{\point}[style=point,estyle=][1]{\,\tikz{\node[style=\commandkey{style}] (x) at (0,-0.05) {$#1$};\draw [\commandkey{estyle}] (x)--(0,1.05);}\,}

\newkeycommand{\copointmap}[style=copoint][1]{\,\tikz{\node[style=\commandkey{style}] (x) at (0,0.3) {$#1$};\draw (x)--(0,-0.7);}\,}

\newkeycommand{\dpoint}[style=point][1]{\,\tikz{\node[style=\commandkey{style},doubled] (x) at (0,-0.3) {$#1$};\draw[boldedge] (x)--(0,0.7);}\,}

\newkeycommand{\kpoint}[style=kpoint,estyle=][1]{\,\tikz{\node[style=\commandkey{style}] (x) at (0,-0.05) {$#1$};\draw [\commandkey{estyle}] (x)--(0,1.05);}\,}

\newkeycommand{\kpointgray}[style=gray kpoint,estyle=][1]{\,\tikz{\node[style=\commandkey{style}] (x) at (0,-0.05) {$#1$};\draw [\commandkey{estyle}] (x)--(0,1.05);}\,}

\newkeycommand{\typedkpoint}[style=kpoint,edgestyle=][2]{%
\,\begin{tikzpicture}
  \begin{pgfonlayer}{nodelayer}
    \node [style=none] (0) at (0, 1) {};
    \node [style=\commandkey{style}] (1) at (0, -0.75) {$#2$};
    \node [style=right label] (2) at (0.25, 0.5) {$#1$};
  \end{pgfonlayer}
  \begin{pgfonlayer}{edgelayer}
    \draw [\commandkey{edgestyle}] (1) to (0.center);
  \end{pgfonlayer}
\end{tikzpicture}\,}

\newkeycommand{\typedkpointdag}[style=kpoint adjoint,edgestyle=][2]{%
\,\begin{tikzpicture}
  \begin{pgfonlayer}{nodelayer}
    \node [style=none] (0) at (0, -1) {};
    \node [style=\commandkey{style}] (1) at (0, 0.75) {$#2$};
    \node [style=right label] (2) at (0.25, -0.5) {$#1$};
  \end{pgfonlayer}
  \begin{pgfonlayer}{edgelayer}
    \draw [\commandkey{edgestyle}] (0.center) to (1);
  \end{pgfonlayer}
\end{tikzpicture}\,}

\newkeycommand{\bistate}[style=kpoint][1]{\,\begin{tikzpicture}
  \begin{pgfonlayer}{nodelayer}
    \node [style=\commandkey{style}, minimum width=1 cm, inner sep=2pt] (0) at (0, -0.25) {$#1$};
    \node [style=none] (1) at (-0.75, 0) {};
    \node [style=none] (2) at (0.75, 0) {};
    \node [style=none] (3) at (-0.75, 0.88) {};
    \node [style=none] (4) at (0.75, 0.88) {};
  \end{pgfonlayer}
  \begin{pgfonlayer}{edgelayer}
    \draw (1.center) to (3.center);
    \draw (2.center) to (4.center);
  \end{pgfonlayer}
\end{tikzpicture}\,}

\newkeycommand{\bistatebig}[style=kpoint][1]{\,\begin{tikzpicture}
  \begin{pgfonlayer}{nodelayer}
    \node [style=\commandkey{style}, minimum width=, minimum width=1.26 cm, inner sep=2pt] (0) at (0, -0.25) {$#1$};
    \node [style=none] (1) at (-1, 0) {};
    \node [style=none] (2) at (1, 0) {};
    \node [style=none] (3) at (-1, 0.88) {};
    \node [style=none] (4) at (1, 0.88) {};
  \end{pgfonlayer}
  \begin{pgfonlayer}{edgelayer}
    \draw (1.center) to (3.center);
    \draw (2.center) to (4.center);
  \end{pgfonlayer}
\end{tikzpicture}\,}

\newkeycommand{\dbistate}[style=dkpoint][1]{\,\begin{tikzpicture}
  \begin{pgfonlayer}{nodelayer}
    \node [style=\commandkey{style}, minimum width=1 cm, inner sep=2pt] (0) at (0, -0.25) {$#1$};
    \node [style=none] (1) at (-0.75, 0) {};
    \node [style=none] (2) at (0.75, 0) {};
    \node [style=none] (3) at (-0.75, 1.0) {};
    \node [style=none] (4) at (0.75, 1.0) {};
  \end{pgfonlayer}
  \begin{pgfonlayer}{edgelayer}
    \draw [boldedge] (1.center) to (3.center);
    \draw [boldedge] (2.center) to (4.center);
  \end{pgfonlayer}
\end{tikzpicture}\,}

\newkeycommand{\bistatebraket}[style1=kpoint,style2=kpointadj][2]{\,\begin{tikzpicture}
  \begin{pgfonlayer}{nodelayer}
    \node [style=\commandkey{style1}, minimum width=1 cm, inner sep=2pt] (0) at (0, -0.75) {$#1$};
    \node [style=none] (1) at (-0.75, -0.5) {};
    \node [style=none] (2) at (0.75, -0.5) {};
    \node [style=none] (3) at (-0.75, 0.5) {};
    \node [style=none] (4) at (0.75, 0.5) {};
    \node [style=\commandkey{style2}, minimum width=1 cm, inner sep=2pt] (5) at (0, 0.75) {$#2$};
  \end{pgfonlayer}
  \begin{pgfonlayer}{edgelayer}
    \draw (1.center) to (3.center);
    \draw (2.center) to (4.center);
  \end{pgfonlayer}
\end{tikzpicture}\,}

\newkeycommand{\weight}[style=dkpoint][1]{\,\begin{tikzpicture}
  \begin{pgfonlayer}{nodelayer}
    \node [style=\commandkey{style}] (0) at (0, -0.5) {$#1$};
    \node [style=upground] (1) at (0, 0.75) {};
  \end{pgfonlayer}
  \begin{pgfonlayer}{edgelayer}
    \draw [style=boldedge] (0) to (1);
  \end{pgfonlayer}
\end{tikzpicture}\,}

\newkeycommand{\dkpoint}[style=dkpoint][1]{\,\tikz{\node[style=\commandkey{style}] (x) at (0,-0.1) {$#1$};\draw[boldedge] (x)--(0,1);}\,}
\newkeycommand{\dkpointadj}[style=dkpointadj][1]{\,\tikz{\node[style=\commandkey{style}] (x) at (0,0.1) {$#1$};\draw[boldedge] (x)--(0,-1);}\,}
\newkeycommand{\dkpointtrans}[style=dkpointtrans][1]{\,\tikz{\node[style=\commandkey{style}] (x) at (0,0.1) {$#1$};\draw[boldedge] (x)--(0,-1);}\,}
\newkeycommand{\dkpointconj}[style=dkpointconj][1]{\,\tikz{\node[style=\commandkey{style}] (x) at (0,-0.1) {$#1$};\draw[boldedge] (x)--(0,1);}\,}

\newkeycommand{\blackkpoint}[style=black kpoint][1]{\,\tikz{\node[style=\commandkey{style}] (x) at (0,-0.1) {$#1$};\draw (x)--(0,1);}\,}
\newkeycommand{\blackkpointadj}[style=black kpointadj][1]{\,\tikz{\node[style=\commandkey{style}] (x) at (0,0.1) {$#1$};\draw (x)--(0,-1);}\,}
\newkeycommand{\blackdkpoint}[style=black dkpoint][1]{\,\tikz{\node[style=\commandkey{style}] (x) at (0,-0.1) {$#1$};\draw[boldedge] (x)--(0,1);}\,}
\newkeycommand{\blackdkpointadj}[style=black dkpointadj][1]{\,\tikz{\node[style=\commandkey{style}] (x) at (0,0.1) {$#1$};\draw[boldedge] (x)--(0,-1);}\,}

\newcommand{\trace}{\,\begin{tikzpicture}
  \begin{pgfonlayer}{nodelayer}
    \node [style=none] (0) at (0, -0.60) {};
    \node [style=upground] (1) at (0, 0.40) {};
  \end{pgfonlayer}
  \begin{pgfonlayer}{edgelayer}
    \draw [style=none] (0.center) to (1);
  \end{pgfonlayer}
\end{tikzpicture}\,}

\newcommand\discard\trace

\newkeycommand{\pointketbra}[style1=copoint,style2=point,estyle=][2]{\,%
\begin{tikzpicture}
  \begin{pgfonlayer}{nodelayer}
    \node [style=none] (0) at (0, -1.75) {};
    \node [style=\commandkey{style1}] (1) at (0, -0.75) {$#1$};
    \node [style=\commandkey{style2}] (2) at (0, 0.75) {$#2$};
    \node [style=none] (3) at (0, 1.75) {};
  \end{pgfonlayer}
  \begin{pgfonlayer}{edgelayer}
    \draw [\commandkey{estyle}] (0.center) to (1);
    \draw [\commandkey{estyle}] (2) to (3.center);
  \end{pgfonlayer}
\end{tikzpicture}\,}

\newkeycommand{\twocopointketbra}[style1=copoint,style2=copoint,style3=point][3]{\,%
\begin{tikzpicture}
  \begin{pgfonlayer}{nodelayer}
    \node [style=none] (0) at (-0.75, -1.75) {};
    \node [style=none] (1) at (0.75, -1.75) {};
    \node [style=\commandkey{style1}] (2) at (-0.75, -0.75) {$#1$};
    \node [style=\commandkey{style2}] (3) at (0.75, -0.75) {$#2$};
    \node [style=\commandkey{style3}] (4) at (0, 0.75) {$#3$};
    \node [style=none] (5) at (0, 1.75) {};
  \end{pgfonlayer}
  \begin{pgfonlayer}{edgelayer}
    \draw (0.center) to (2);
    \draw (1.center) to (3);
    \draw (4) to (5.center);
  \end{pgfonlayer}
\end{tikzpicture}\,}

\newkeycommand{\twopointketbra}[style1=copoint,style2=point,style3=point][3]{\,%
\begin{tikzpicture}
  \begin{pgfonlayer}{nodelayer}
    \node [style=none] (0) at (0, -1.75) {};
    \node [style=none] (1) at (-0.75, 1.75) {};
    \node [style=\commandkey{style1}] (2) at (0, -0.75) {$#1$};
    \node [style=\commandkey{style2}] (3) at (-0.75, 0.75) {$#2$};
    \node [style=\commandkey{style3}] (4) at (0.75, 0.75) {$#3$};
    \node [style=none] (5) at (0.75, 1.75) {};
  \end{pgfonlayer}
  \begin{pgfonlayer}{edgelayer}
    \draw (0.center) to (2);
    \draw (1.center) to (3);
    \draw (4) to (5.center);
  \end{pgfonlayer}
\end{tikzpicture}\,}

\newkeycommand{\pointbraket}[style1=point,style2=copoint,estyle=][2]{\,%
\begin{tikzpicture}
  \begin{pgfonlayer}{nodelayer}
    \node [style=\commandkey{style1}] (0) at (0, -0.625) {$#1$};
    \node [style=\commandkey{style2}] (1) at (0, 0.625) {$#2$};
  \end{pgfonlayer}
  \begin{pgfonlayer}{edgelayer}
    \draw [\commandkey{estyle}] (0) to (1);
  \end{pgfonlayer}
\end{tikzpicture}\,}

\newkeycommand{\kpointketbra}[style1=kpointdag,style2=kpoint,estyle=][2]{\,%
\begin{tikzpicture}
  \begin{pgfonlayer}{nodelayer}
    \node [style=none] (0) at (0, -1.9) {};
    \node [style=\commandkey{style1}] (1) at (0, -0.95) {$#1$};
    \node [style=\commandkey{style2}] (2) at (0, 0.95) {$#2$};
    \node [style=none] (3) at (0, 1.9) {};
  \end{pgfonlayer}
  \begin{pgfonlayer}{edgelayer}
    \draw [\commandkey{estyle}] (0.center) to (1);
    \draw [\commandkey{estyle}] (2) to (3.center);
  \end{pgfonlayer}
\end{tikzpicture}\,}

\newkeycommand{\kpointmap}[style1=kpoint,style2=map][2]{\,%
\begin{tikzpicture}
  \begin{pgfonlayer}{nodelayer}
    \node [style=\commandkey{style1}] (0) at (0, -1.1) {$#1$};
    \node [style=\commandkey{style2}] (1) at (0, 0.5) {$#2$};
    \node [style=none] (2) at (0, 1.75) {};
  \end{pgfonlayer}
  \begin{pgfonlayer}{edgelayer}
    \draw (0) to (1);
    \draw (1) to (2.center);
  \end{pgfonlayer}
\end{tikzpicture}%
\,}

\newkeycommand{\kpointmaptrans}[style1=kpointconj,style2=mapadj][2]{\,%
\begin{tikzpicture}
  \begin{pgfonlayer}{nodelayer}
    \node [style=\commandkey{style1}] (0) at (0, -1.1) {$#1$};
    \node [style=\commandkey{style2}] (1) at (0, 0.5) {$#2$};
    \node [style=none] (2) at (0, 1.75) {};
  \end{pgfonlayer}
  \begin{pgfonlayer}{edgelayer}
    \draw (0) to (1);
    \draw (1) to (2.center);
  \end{pgfonlayer}
\end{tikzpicture}%
\,}

\newkeycommand{\kpointmapadj}[style1=kpointadj,style2=map][2]{\,%
\begin{tikzpicture}
  \begin{pgfonlayer}{nodelayer}
    \node [style=\commandkey{style1}] (0) at (0, 1.1) {$#1$};
    \node [style=\commandkey{style2}] (1) at (0, -0.5) {$#2$};
    \node [style=none] (2) at (0, -1.75) {};
  \end{pgfonlayer}
  \begin{pgfonlayer}{edgelayer}
    \draw (0) to (1);
    \draw (1) to (2.center);
  \end{pgfonlayer}
\end{tikzpicture}%
\,}

\newkeycommand{\dkpointmap}[style1=dkpoint,style2=dmap][2]{\,%
\begin{tikzpicture}
  \begin{pgfonlayer}{nodelayer}
    \node [style=\commandkey{style1}] (0) at (0, -1.2) {$#1$};
    \node [style=\commandkey{style2}] (1) at (0, 0.4) {$#2$};
    \node [style=none] (2) at (0, 1.7) {};
  \end{pgfonlayer}
  \begin{pgfonlayer}{edgelayer}
    \draw[style=boldedge] (0) to (1);
    \draw[style=boldedge] (1) to (2.center);
  \end{pgfonlayer}
\end{tikzpicture}%
\,}

\newkeycommand{\dkpointmapx}[style1=dkpoint,style2=dmap][2]{\,%
\begin{tikzpicture}
  \begin{pgfonlayer}{nodelayer}
    \node [style=\commandkey{style1}] (0) at (0, -1.4) {$#1$};
    \node [style=\commandkey{style2}] (1) at (0, 0.6) {$#2$};
    \node [style=none] (2) at (0, 1.9) {};
  \end{pgfonlayer}
  \begin{pgfonlayer}{edgelayer}
    \draw[style=boldedge] (0) to (1);
    \draw[style=boldedge] (1) to (2.center);
  \end{pgfonlayer}
\end{tikzpicture}%
\,}

\newkeycommand{\boxcopointmapdag}[style1=map,style2=copoint][2]{\,%
\begin{tikzpicture}
  \begin{pgfonlayer}{nodelayer}
    \node [style=none] (0) at (0, -1.75) {};
    \node [style=\commandkey{style1}] (1) at (0, -0.5) {$#1$};
    \node [style=\commandkey{style2}] (2) at (0, 1) {$#2$};
  \end{pgfonlayer}
  \begin{pgfonlayer}{edgelayer}
    \draw (0.center) to (1);
    \draw (1) to (2);
  \end{pgfonlayer}
\end{tikzpicture}%
\,}

\newkeycommand{\kpointdagmap}[style1=map,style2=kpointdag][2]{\,%
\begin{tikzpicture}
  \begin{pgfonlayer}{nodelayer}
    \node [style=none] (0) at (0, -1.75) {};
    \node [style=\commandkey{style1}] (1) at (0, -0.5) {$#1$};
    \node [style=\commandkey{style2}] (2) at (0, 1.1) {$#2$};
  \end{pgfonlayer}
  \begin{pgfonlayer}{edgelayer}
    \draw (0.center) to (1);
    \draw (1) to (2);
  \end{pgfonlayer}
\end{tikzpicture}%
\,}

\newkeycommand{\sandwichmap}[style1=point,style2=map,style3=copoint][3]{\,%
\begin{tikzpicture}
  \begin{pgfonlayer}{nodelayer}
    \node [style=\commandkey{style1}] (0) at (0, -1.50) {$#1$};
    \node [style=\commandkey{style2}] (1) at (0, 0) {$#2$};
    \node [style=\commandkey{style3}] (2) at (0, 1.50) {$#3$};
  \end{pgfonlayer}
  \begin{pgfonlayer}{edgelayer}
    \draw (0) to (1);
    \draw (1) to (2);
  \end{pgfonlayer}
\end{tikzpicture}%
}

\newkeycommand{\kpointsandwichmap}[style1=kpoint,style2=map,style3=kpointdag][3]{\,%
\begin{tikzpicture}
  \begin{pgfonlayer}{nodelayer}
    \node [style=\commandkey{style1}] (0) at (0, -1.5) {$#1$};
    \node [style=\commandkey{style2}] (1) at (0, 0) {$#2$};
    \node [style=\commandkey{style3}] (2) at (0, 1.5) {$#3$};
  \end{pgfonlayer}
  \begin{pgfonlayer}{edgelayer}
    \draw (0) to (1);
    \draw (1) to (2);
  \end{pgfonlayer}
\end{tikzpicture}%
}

\newkeycommand{\sandwichmapconj}[style1=point,style2=mapconj,style3=copoint][3]{\,% 
\begin{tikzpicture}
  \begin{pgfonlayer}{nodelayer}
    \node [style=\commandkey{style1}] (0) at (0, -1.50) {$#1$};
    \node [style=\commandkey{style2}] (1) at (0, 0) {$#2$};
    \node [style=\commandkey{style3}] (2) at (0, 1.50) {$#3$};
  \end{pgfonlayer}
  \begin{pgfonlayer}{edgelayer}
    \draw (0) to (1);
    \draw (1) to (2);
  \end{pgfonlayer}
\end{tikzpicture}%
}

\newkeycommand{\sandwichtwo}[style1=point,style2=map,style3=map,style4=copoint][4]{\,%
\begin{tikzpicture}
  \begin{pgfonlayer}{nodelayer}
    \node [style=\commandkey{style2}] (0) at (0, -1) {$#2$};
    \node [style=\commandkey{style3}] (1) at (0, 1) {$#3$};
    \node [style=\commandkey{style1}] (2) at (0, -2.6) {$#1$};
    \node [style=\commandkey{style4}] (3) at (0, 2.6) {$#4$};
  \end{pgfonlayer}
  \begin{pgfonlayer}{edgelayer}
    \draw (2) to (0);
    \draw (0) to (1);
    \draw (1) to (3);
  \end{pgfonlayer}
\end{tikzpicture}\,}

\newkeycommand{\longbraket}[style1=point,style2=copoint][2]{\,% 
\begin{tikzpicture}
  \begin{pgfonlayer}{nodelayer}
    \node [style=\commandkey{style1}] (0) at (0, -2.6) {$#1$};
    \node [style=\commandkey{style2}] (1) at (0, 2.6) {$#2$};
  \end{pgfonlayer}
  \begin{pgfonlayer}{edgelayer}
    \draw (0) to (1);
  \end{pgfonlayer}
\end{tikzpicture}\,}

\newkeycommand{\onb}[style=point]{\ensuremath{\{\,\tikz{\node[style=\commandkey{style}] (x) at (0,-0.3) {$j$};\draw (x)--(0,0.7);}\,\}}\xspace}

\newkeycommand{\phase}[style=white phase dot][1]{\,\begin{tikzpicture}
    \begin{pgfonlayer}{nodelayer}
        \node [style=none] (0) at (0, 1) {};
        \node [style=\commandkey{style}] (2) at (0, -0) {$#1$}; 
        \node [style=none] (3) at (0, -1) {};
    \end{pgfonlayer}
    \begin{pgfonlayer}{edgelayer}
        \draw (2) to (0.center);
        \draw (3.center) to (2);
    \end{pgfonlayer}
\end{tikzpicture}\,}

\newkeycommand{\dphase}[style=white phase ddot][1]{\,\begin{tikzpicture}
    \begin{pgfonlayer}{nodelayer}
        \node [style=none] (0) at (0, 1.25) {};
        \node [style=\commandkey{style}] (2) at (0, -0) {$#1$};
        \node [style=none] (3) at (0, -1.25) {};
    \end{pgfonlayer}
    \begin{pgfonlayer}{edgelayer}
        \draw [boldedge] (2) to (0.center);
        \draw [boldedge] (3.center) to (2);
    \end{pgfonlayer}
\end{tikzpicture}\,}

\newkeycommand{\dphasepoint}[style=white phase ddot][1]{\,\begin{tikzpicture}[yshift=-3mm]
  \begin{pgfonlayer}{nodelayer}
    \node [style=none] (0) at (0, 1) {};
    \node [style=\commandkey{style}] (1) at (0, 0) {$#1$};
  \end{pgfonlayer}
  \begin{pgfonlayer}{edgelayer}
    \draw [style=boldedge] (0.center) to (1);
  \end{pgfonlayer}
\end{tikzpicture}\,}

\newkeycommand{\dphasepointgray}[style=gray phase ddot][1]{\,\begin{tikzpicture}[yshift=-3mm]
  \begin{pgfonlayer}{nodelayer}
    \node [style=none] (0) at (0, 1) {};
    \node [style=\commandkey{style}] (1) at (0, 0) {$#1$};
  \end{pgfonlayer}
  \begin{pgfonlayer}{edgelayer}
    \draw [style=boldedge] (0.center) to (1);
  \end{pgfonlayer}
\end{tikzpicture}\,}

\newkeycommand{\dphasecopoint}[style=white phase ddot][1]{\,\begin{tikzpicture}[yshift=3mm,yscale=-1]
  \begin{pgfonlayer}{nodelayer}
    \node [style=none] (0) at (0, 1) {};
    \node [style=\commandkey{style}] (1) at (0, 0) {$#1$};
  \end{pgfonlayer}
  \begin{pgfonlayer}{edgelayer}
    \draw [style=boldedge] (0.center) to (1);
  \end{pgfonlayer}
\end{tikzpicture}\,}

\newkeycommand{\dphasepointsm}[style=white phase ddot][1]{\,\begin{tikzpicture}[yshift=-3mm]
  \begin{pgfonlayer}{nodelayer}
    \node [style=none] (0) at (0, 1) {};
    \node [style=\commandkey{style}] (1) at (0, 0) {\footnotesize\!$#1$\!};
  \end{pgfonlayer}
  \begin{pgfonlayer}{edgelayer}
    \draw [style=boldedge] (0.center) to (1);
  \end{pgfonlayer}
\end{tikzpicture}\,}

\newkeycommand{\phasepoint}[style=white phase dot][1]{\,\begin{tikzpicture}[yshift=-3mm]
  \begin{pgfonlayer}{nodelayer}
    \node [style=none] (0) at (0, 1) {};
    \node [style=\commandkey{style}] (1) at (0, 0) {$#1$};
  \end{pgfonlayer}
  \begin{pgfonlayer}{edgelayer}
    \draw (0.center) to (1);
  \end{pgfonlayer}
\end{tikzpicture}\,}

\newkeycommand{\phasecopoint}[style=white phase dot][1]{\,\begin{tikzpicture}[yshift=3mm]
  \begin{pgfonlayer}{nodelayer}
    \node [style=none] (0) at (0, -1) {};
    \node [style=\commandkey{style}] (1) at (0, 0) {$#1$};
  \end{pgfonlayer}
  \begin{pgfonlayer}{edgelayer}
    \draw (0.center) to (1);
  \end{pgfonlayer}
\end{tikzpicture}\,}

\newkeycommand{\kpointbraket}[style1=kpoint,style2=kpoint adjoint,estyle=][2]{\,%
\begin{tikzpicture}
  \begin{pgfonlayer}{nodelayer}
    \node [style=\commandkey{style1}] (0) at (0, -0.735) {$#1$};
    \node [style=\commandkey{style2}] (1) at (0, 0.735) {$#2$};
  \end{pgfonlayer}
  \begin{pgfonlayer}{edgelayer}
    \draw [\commandkey{estyle}] (0) to (1);
  \end{pgfonlayer}
\end{tikzpicture}\,}

\newkeycommand{\kkpointbraket}[style1=kpoint,style2=kpoint adjoint,estyle=][2]{\,%
\begin{tikzpicture}
  \begin{pgfonlayer}{nodelayer}
    \node [style=\commandkey{style1}] (0) at (0, -0.685) {$#1$};
    \node [style=\commandkey{style2}] (1) at (0, 0.685) {$#2$};
  \end{pgfonlayer}
  \begin{pgfonlayer}{edgelayer}
    \draw [\commandkey{estyle}] (0) to (1);
  \end{pgfonlayer}
\end{tikzpicture}\,}

\newkeycommand{\kpointbraketx}[style1=kpoint,style2=kpoint adjoint,estyle=][2]{\,%
\begin{tikzpicture}
  \begin{pgfonlayer}{nodelayer}
    \node [style=\commandkey{style1}] (0) at (0, -0.885) {$#1$};
    \node [style=\commandkey{style2}] (1) at (0, 0.885) {$#2$};
  \end{pgfonlayer}
  \begin{pgfonlayer}{edgelayer}
    \draw [\commandkey{estyle}] (0) to (1);
  \end{pgfonlayer}
\end{tikzpicture}\,}

\tikzstyle{cdiag}=[matrix of math nodes, row sep=3em, column sep=3em, text height=1.5ex, text depth=0.25ex,inner sep=0.5em]
\tikzstyle{arrow above}=[transform canvas={yshift=0.5ex}]
\tikzstyle{arrow below}=[transform canvas={yshift=-0.5ex}]

\usepackage{tikz}
\usetikzlibrary{decorations.markings}
\usetikzlibrary{shapes.geometric}

\pgfdeclarelayer{edgelayer}
\pgfdeclarelayer{nodelayer}
\pgfsetlayers{edgelayer,nodelayer,main}

\tikzstyle{none}=[inner sep=0pt]
\definecolor{hexcolor0xff0000}{rgb}{1.000,0.000,0.000}
\definecolor{hexcolor0x000000}{rgb}{0.000,0.000,0.000}
\definecolor{hexcolor0x00ff00}{rgb}{0.000,1.000,0.000}
\definecolor{hexcolor0x000000}{rgb}{0.000,0.000,0.000}
\definecolor{hexcolor0xffff00}{rgb}{1.000,1.000,0.000}
\definecolor{hexcolor0xffffff}{rgb}{1.000,1.000,1.000}

\tikzstyle{rn}=[circle,fill=hexcolor0xff0000,draw=hexcolor0x000000,line width=0.8 pt]
\tikzstyle{gn}=[circle,fill=hexcolor0x00ff00,draw=hexcolor0x000000,line width=0.8 pt]
\tikzstyle{yn}=[circle,fill=hexcolor0xffff00,draw=hexcolor0x000000,line width=0.8 pt]
\tikzstyle{wn}=[circle,fill=hexcolor0xffffff,draw=hexcolor0x000000,line width=0.8 pt]
\tikzstyle{wnthick}=[circle,fill=hexcolor0xffffff,draw=hexcolor0x000000,line width=2.500]

\tikzstyle{simple}=[-,draw=hexcolor0x000000,line width=2.000]
\tikzstyle{arrow}=[-,draw=hexcolor0x000000,postaction={decorate},decoration={markings,mark=at position .5 with {\arrow{>}}},line width=2.000]
\tikzstyle{tick}=[-,draw=hexcolor0x000000,postaction={decorate},decoration={markings,mark=at position .5 with {\draw (0,-0.1) -- (0,0.1);}},line width=2.000]
\tikzstyle{halfthickness}=[-,draw=hexcolor0x000000,line width=0.500]
\tikzstyle{thick}=[-,draw=hexcolor0x000000,line width=2.500]
\tikzstyle{thicker}=[-,draw=hexcolor0x000000,line width=4.000]

\tikzstyle{env}=[copoint,regular polygon rotate=0,minimum width=0.2cm, fill=black]

\tikzstyle{probs}=[shape=semicircle,fill=white,draw=black,shape border rotate=180,minimum width=1.2cm]

\tikzstyle{every picture}=[baseline=-0.25em,scale=0.5]
\tikzstyle{dotpic}=[] % for backwards-compatibility
\tikzstyle{diredges}=[every to/.style={diredge}]
\tikzstyle{math matrix}=[matrix of math nodes,left delimiter=(,right delimiter=),inner sep=2pt,column sep=1em,row sep=0.5em,nodes={inner sep=0pt},text height=1.5ex, text depth=0.25ex]

\tikzstyle{inline text}=[text height=1.5ex, text depth=0.25ex,yshift=0.5mm]
\tikzstyle{label}=[font=\footnotesize,text height=1.5ex, text depth=0.25ex,yshift=0.5mm]
\tikzstyle{left label}=[label,anchor=east,xshift=1.5mm]
\tikzstyle{right label}=[label,anchor=west,xshift=-1.5mm]

\tikzstyle{braceedge}=[decorate,decoration={brace,amplitude=2mm,raise=-1mm}]
\tikzstyle{small braceedge}=[decorate,decoration={brace,amplitude=1mm,raise=-1mm}]

\tikzstyle{doubled}=[line width=1.6pt] % set the line width for all doubled (quantum) maps/wires
\tikzstyle{boldedge}=[doubled,shorten <=-0.17mm,shorten >=-0.17mm]
\tikzstyle{boldedgegray}=[doubled,gray,shorten <=-0.17mm,shorten >=-0.17mm]

\tikzstyle{semidoubled}=[line width=1.4pt] % set the line width for all doubled (quantum) maps/wires
\tikzstyle{semiboldedgegray}=[semidoubled,gray,shorten <=-0.17mm,shorten >=-0.17mm]

\tikzstyle{boldedgedashed}=[very thick,dashed,shorten <=-0.17mm,shorten >=-0.17mm]
\tikzstyle{vboldedgedashed}=[doubled,dashed,shorten <=-0.17mm,shorten >=-0.17mm]
\tikzstyle{left hook arrow}=[left hook-latex]
\tikzstyle{right hook arrow}=[right hook-latex]
\tikzstyle{sembracket}=[line width=0.5pt,shorten <=-0.07mm,shorten >=-0.07mm]

\tikzstyle{causal edge}=[->,thick,gray]
\tikzstyle{causal nondir}=[thick,gray]
\tikzstyle{timeline}=[thick,gray, dashed]

\tikzstyle{cedge}=[<->,thick,gray!70!white]

\tikzstyle{empty diagram}=[draw=gray!40!white,dashed,shape=rectangle,minimum width=1cm,minimum height=1cm]
\tikzstyle{empty diagram small}=[draw=gray!50!white,dashed,shape=rectangle,minimum width=0.6cm,minimum height=0.5cm]

\tikzstyle{dot}=[inner sep=0mm,minimum width=2mm,minimum height=2mm,draw,shape=circle]
\tikzstyle{ddot}=[inner sep=0mm, doubled, minimum width=2.5mm,minimum height=2.5mm,draw,shape=circle]

\tikzstyle{black dot}=[dot,fill=black]
\tikzstyle{white dot}=[dot,fill=white,,text depth=-0.2mm]
\tikzstyle{green dot}=[white dot] % for backwards-compatibility
\tikzstyle{gray dot}=[dot,fill=gray!40!white,,text depth=-0.2mm]
\tikzstyle{red dot}=[gray dot] % for backwards-compatibility

\tikzstyle{black ddot}=[ddot,fill=black]
\tikzstyle{white ddot}=[ddot,fill=white]
\tikzstyle{gray ddot}=[ddot,fill=gray!40!white]

\tikzstyle{gray edge}=[gray!40!white]

\tikzstyle{small dot}=[inner sep=0.5mm,minimum width=0pt,minimum height=0pt,draw,shape=circle]

\tikzstyle{small black dot}=[small dot,fill=black]
\tikzstyle{small white dot}=[small dot,fill=white]
\tikzstyle{small gray dot}=[small dot,fill=gray!40!white]

\tikzstyle{causal dot}=[inner sep=0.4mm,minimum width=0pt,minimum height=0pt,draw=white,shape=circle,fill=gray!40!white]

\tikzstyle{phase dimensions}=[minimum size=5mm,font=\footnotesize,rectangle,rounded corners=2.5mm,inner sep=0.2mm,outer sep=-2mm]
\tikzstyle{phase dimensions small}=[minimum size=3.0mm,font=\footnotesize,rectangle,rounded corners=1.5mm,inner sep=0.2mm,outer sep=-1.2mm]
\tikzstyle{dphase dimensions}=[minimum size=5mm,font=\footnotesize,rectangle,rounded corners=2.5mm,inner sep=0.2mm,outer sep=-2mm]
\tikzstyle{white phase dot}=[dot,fill=white,phase dimensions]
\tikzstyle{white phase dot small}=[dot,fill=white,phase dimensions small]
\tikzstyle{white phase ddot}=[ddot,fill=white,dphase dimensions]
\tikzstyle{green phase ddot}=[ddot,fill=green,dphase dimensions]

\tikzstyle{white rect ddot}=[draw=black,fill=white,doubled,minimum size=5mm,font=\footnotesize,rectangle,rounded corners=2.5mm,inner sep=0.2mm]
\tikzstyle{gray rect ddot}=[draw=black,fill=gray!40!white,doubled,minimum size=6mm,font=\footnotesize,rectangle,rounded corners=3mm]

\tikzstyle{gray phase dot}=[dot,fill=gray!40!white,phase dimensions]
\tikzstyle{gray phase dot small}=[dot,fill=gray!40!white,phase dimensions small]
\tikzstyle{gray phase ddot}=[ddot,fill=gray!40!white,dphase dimensions]
\tikzstyle{red phase ddot}=[ddot,fill=red,dphase dimensions]

\tikzstyle{grey phase dot}=[gray phase dot]
\tikzstyle{grey phase ddot}=[gray phase ddot]

\tikzstyle{small phase dimensions}=[minimum size=4mm,font=\tiny,rectangle,rounded corners=2mm,inner sep=0.2mm,outer sep=-2mm]
\tikzstyle{small dphase dimensions}=[minimum size=4mm,font=\tiny,rectangle,rounded corners=2mm,inner sep=0.2mm,outer sep=-2mm]

\tikzstyle{small gray phase dot}=[dot,fill=gray!40!white,small phase dimensions]
\tikzstyle{small gray phase ddot}=[ddot,fill=gray!40!white,small dphase dimensions]

\tikzstyle{small map}=[draw,shape=rectangle,minimum height=4mm,minimum width=4mm,fill=white]

\tikzstyle{cnot}=[fill=white,shape=circle,inner sep=-1.4pt]

\tikzstyle{asym hadamard}=[fill=white,draw,shape=NEbox,inner sep=0.6mm,font=\footnotesize,minimum height=4mm]
\tikzstyle{asym hadamard conj}=[fill=white,draw,shape=NWbox,inner sep=0.6mm,font=\footnotesize,minimum height=4mm]
\tikzstyle{asym hadamard dag}=[fill=white,draw,shape=SEbox,inner sep=0.6mm,font=\footnotesize,minimum height=4mm]

\tikzstyle{hadamard}=[fill=white,draw,inner sep=0.6mm,font=\footnotesize,minimum height=4mm,minimum width=4mm]
\tikzstyle{small hadamard}=[fill=white,draw,inner sep=0.6mm,minimum height=1.5mm,minimum width=1.5mm]
\tikzstyle{dhadamard}=[hadamard,doubled]
\tikzstyle{small dhadamard}=[small hadamard,doubled]
\tikzstyle{small dhadamard rotate}=[small hadamard,doubled,rotate=45]
\tikzstyle{antipode}=[white dot,inner sep=0.3mm,font=\footnotesize]

\tikzstyle{scalar}=[diamond,draw,inner sep=0.5pt,font=\small]
\tikzstyle{dscalar}=[diamond,doubled, draw,inner sep=0.5pt,font=\small]

\tikzstyle{small box}=[rectangle,inline text,fill=white,draw,minimum height=5mm,yshift=-0.5mm,minimum width=5mm,font=\small]
\tikzstyle{small gray box}=[small box,fill=gray!30]
\tikzstyle{medium box}=[rectangle,inline text,fill=white,draw,minimum height=5mm,yshift=-0.5mm,minimum width=10mm,font=\small]
\tikzstyle{square box}=[small box] % for backwards-compatibility
\tikzstyle{medium gray box}=[small box,fill=gray!30]
\tikzstyle{semilarge box}=[rectangle,inline text,fill=white,draw,minimum height=5mm,yshift=-0.5mm,minimum width=12.5mm,font=\small]
\tikzstyle{large box}=[rectangle,inline text,fill=white,draw,minimum height=5mm,yshift=-0.5mm,minimum width=15mm,font=\small]
\tikzstyle{large gray box}=[small box,fill=gray!30]

\tikzstyle{Bayes box}=[rectangle,fill=black,draw, minimum height=3mm, minimum width=3mm]

\tikzstyle{gray square point}=[small box,fill=gray!50]

\tikzstyle{dphase box white}=[dhadamard]
\tikzstyle{dphase box gray}=[dhadamard,fill=gray!50!white]

\tikzstyle{point}=[regular polygon,regular polygon sides=3,draw,scale=0.75,inner sep=-0.5pt,minimum width=9mm,fill=white,regular polygon rotate=180]
\tikzstyle{copoint}=[regular polygon,regular polygon sides=3,draw,scale=0.75,inner sep=-0.5pt,minimum width=9mm,fill=white]
\tikzstyle{dpoint}=[point,doubled]
\tikzstyle{dcopoint}=[copoint,doubled]

\tikzstyle{wide copoint}=[fill=white,draw,shape=isosceles triangle,shape border rotate=90,isosceles triangle stretches=true,inner sep=0pt,minimum width=1.5cm,minimum height=6.12mm]
\tikzstyle{wide point}=[fill=white,draw,shape=isosceles triangle,shape border rotate=-90,isosceles triangle stretches=true,inner sep=0pt,minimum width=1.5cm,minimum height=6.12mm,yshift=-0.0mm]
\tikzstyle{wide point plus}=[fill=white,draw,shape=isosceles triangle,shape border rotate=-90,isosceles triangle stretches=true,inner sep=0pt,minimum width=1.74cm,minimum height=7mm,yshift=-0.0mm]

\tikzstyle{wide dpoint}=[fill=white,doubled,draw,shape=isosceles triangle,shape border rotate=-90,isosceles triangle stretches=true,inner sep=0pt,minimum width=1.5cm,minimum height=6.12mm,yshift=-0.0mm]
\tikzstyle{wide dcopoint}=[fill=white,doubled,draw,shape=isosceles triangle,shape border rotate=90,isosceles triangle stretches=true,inner sep=0pt,minimum width=1.5cm,minimum height=6.12mm,yshift=-0.0mm]

\tikzstyle{tinypoint}=[regular polygon,regular polygon sides=3,draw,scale=0.55,inner sep=-0.15pt,minimum width=6mm,fill=white,regular polygon rotate=180]

\tikzstyle{white point}=[point]
\tikzstyle{white dpoint}=[dpoint]
\tikzstyle{green point}=[white point] % for backwards-compatibility
\tikzstyle{white copoint}=[copoint]
\tikzstyle{gray point}=[point,fill=gray!40!white]
\tikzstyle{gray dpoint}=[gray point,doubled]
\tikzstyle{red point}=[gray point] % for backwards-compatibility
\tikzstyle{gray copoint}=[copoint,fill=gray!40!white]
\tikzstyle{gray dcopoint}=[gray copoint,doubled]

\tikzstyle{white point guide}=[regular polygon,regular polygon sides=3,font=\scriptsize,draw,scale=0.65,inner sep=-0.5pt,minimum width=9mm,fill=white,regular polygon rotate=180]

\tikzstyle{black point}=[point,fill=black,font=\color{white}]
\tikzstyle{black copoint}=[copoint,fill=black,font=\color{white}]

\tikzstyle{tiny gray point}=[tinypoint,fill=gray!40!white]

\tikzstyle{diredge}=[->]
\tikzstyle{ddiredge}=[<->]
\tikzstyle{rdiredge}=[<-]
\tikzstyle{thickdiredge}=[->, very thick]
\tikzstyle{pointer edge}=[->,very thick,gray]
\tikzstyle{pointer edge part}=[very thick,gray]
\tikzstyle{dashed edge}=[dashed]
\tikzstyle{thick dashed edge}=[very thick,dashed]
\tikzstyle{thick gray dashed edge}=[thick dashed edge,gray!40]
\tikzstyle{thick map edge}=[very thick,|->]

\makeatletter
\newcommand{\boxshape}[3]{%
\pgfdeclareshape{#1}{
\inheritsavedanchors[from=rectangle] % this is nearly a rectangle
\inheritanchorborder[from=rectangle]
\inheritanchor[from=rectangle]{center}
\inheritanchor[from=rectangle]{north}
\inheritanchor[from=rectangle]{south}
\inheritanchor[from=rectangle]{west}
\inheritanchor[from=rectangle]{east}
\backgroundpath{% this is new
\southwest \pgf@xa=\pgf@x \pgf@ya=\pgf@y
\northeast \pgf@xb=\pgf@x \pgf@yb=\pgf@y

\@tempdima=#2
\@tempdimb=#3

\pgfpathmoveto{\pgfpoint{\pgf@xa - 5pt + \@tempdima}{\pgf@ya}}
\pgfpathlineto{\pgfpoint{\pgf@xa - 5pt - \@tempdima}{\pgf@yb}}
\pgfpathlineto{\pgfpoint{\pgf@xb + 5pt + \@tempdimb}{\pgf@yb}}
\pgfpathlineto{\pgfpoint{\pgf@xb + 5pt - \@tempdimb}{\pgf@ya}}
\pgfpathlineto{\pgfpoint{\pgf@xa - 5pt + \@tempdima}{\pgf@ya}}
\pgfpathclose
}
}}

\boxshape{NEbox}{0pt}{5pt}
\boxshape{SEbox}{0pt}{-5pt}
\boxshape{NWbox}{5pt}{0pt}
\boxshape{SWbox}{-5pt}{0pt}
\boxshape{EBox}{-3pt}{3pt}
\boxshape{WBox}{3pt}{-3pt}
\makeatother

\tikzstyle{cloud}=[shape=cloud,draw,minimum width=1.5cm,minimum height=1.5cm]

\tikzstyle{map}=[draw,shape=NEbox,inner sep=2pt,minimum height=6mm,fill=white]
\tikzstyle{dashedmap}=[draw,dashed,shape=NEbox,inner sep=2pt,minimum height=6mm,fill=white]
\tikzstyle{mapdag}=[draw,shape=SEbox,inner sep=2pt,minimum height=6mm,fill=white]
\tikzstyle{mapadj}=[draw,shape=SEbox,inner sep=2pt,minimum height=6mm,fill=white]
\tikzstyle{maptrans}=[draw,shape=SWbox,inner sep=2pt,minimum height=6mm,fill=white]
\tikzstyle{mapconj}=[draw,shape=NWbox,inner sep=2pt,minimum height=6mm,fill=white]

\tikzstyle{medium map}=[draw,shape=NEbox,inner sep=2pt,minimum height=6mm,fill=white,minimum width=7mm]
\tikzstyle{medium map dag}=[draw,shape=SEbox,inner sep=2pt,minimum height=6mm,fill=white,minimum width=7mm]
\tikzstyle{medium map adj}=[draw,shape=SEbox,inner sep=2pt,minimum height=6mm,fill=white,minimum width=7mm]
\tikzstyle{medium map trans}=[draw,shape=SWbox,inner sep=2pt,minimum height=6mm,fill=white,minimum width=7mm]
\tikzstyle{medium map conj}=[draw,shape=NWbox,inner sep=2pt,minimum height=6mm,fill=white,minimum width=7mm]
\tikzstyle{semilarge map}=[draw,shape=NEbox,inner sep=2pt,minimum height=6mm,fill=white,minimum width=9.5mm]
\tikzstyle{semilarge map trans}=[draw,shape=SWbox,inner sep=2pt,minimum height=6mm,fill=white,minimum width=9.5mm]
\tikzstyle{semilarge map adj}=[draw,shape=SEbox,inner sep=2pt,minimum height=6mm,fill=white,minimum width=9.5mm]
\tikzstyle{semilarge map dag}=[draw,shape=SEbox,inner sep=2pt,minimum height=6mm,fill=white,minimum width=9.5mm]
\tikzstyle{semilarge map conj}=[draw,shape=NWbox,inner sep=2pt,minimum height=6mm,fill=white,minimum width=9.5mm]
\tikzstyle{large map}=[draw,shape=NEbox,inner sep=2pt,minimum height=6mm,fill=white,minimum width=12mm]
\tikzstyle{large map conj}=[draw,shape=NWbox,inner sep=2pt,minimum height=6mm,fill=white,minimum width=12mm]
\tikzstyle{very large map}=[draw,shape=NEbox,inner sep=2pt,minimum height=6mm,fill=white,minimum width=17mm]

\tikzstyle{medium dmap}=[draw,doubled,shape=NEbox,inner sep=2pt,minimum height=6mm,fill=white,minimum width=7mm]
\tikzstyle{medium dmap dag}=[draw,doubled,shape=SEbox,inner sep=2pt,minimum height=6mm,fill=white,minimum width=7mm]
\tikzstyle{medium dmap adj}=[draw,doubled,shape=SEbox,inner sep=2pt,minimum height=6mm,fill=white,minimum width=7mm]
\tikzstyle{medium dmap trans}=[draw,doubled,shape=SWbox,inner sep=2pt,minimum height=6mm,fill=white,minimum width=7mm]
\tikzstyle{medium dmap conj}=[draw,doubled,shape=NWbox,inner sep=2pt,minimum height=6mm,fill=white,minimum width=7mm]
\tikzstyle{semilarge dmap}=[draw,doubled,shape=NEbox,inner sep=2pt,minimum height=6mm,fill=white,minimum width=9.5mm]
\tikzstyle{semilarge dmap trans}=[draw,doubled,shape=SWbox,inner sep=2pt,minimum height=6mm,fill=white,minimum width=9.5mm]
\tikzstyle{semilarge dmap adj}=[draw,doubled,shape=SEbox,inner sep=2pt,minimum height=6mm,fill=white,minimum width=9.5mm]
\tikzstyle{semilarge dmap dag}=[draw,doubled,shape=SEbox,inner sep=2pt,minimum height=6mm,fill=white,minimum width=9.5mm]
\tikzstyle{semilarge dmap conj}=[draw,doubled,shape=NWbox,inner sep=2pt,minimum height=6mm,fill=white,minimum width=9.5mm]
\tikzstyle{large dmap}=[draw,doubled,shape=NEbox,inner sep=2pt,minimum height=6mm,fill=white,minimum width=12mm]
\tikzstyle{large dmap conj}=[draw,doubled,shape=NWbox,inner sep=2pt,minimum height=6mm,fill=white,minimum width=12mm]
\tikzstyle{large dmap trans}=[draw,doubled,shape=SWbox,inner sep=2pt,minimum height=6mm,fill=white,minimum width=12mm]
\tikzstyle{large dmap adj}=[draw,doubled,shape=SEbox,inner sep=2pt,minimum height=6mm,fill=white,minimum width=12mm]
\tikzstyle{large dmap dag}=[draw,doubled,shape=SEbox,inner sep=2pt,minimum height=6mm,fill=white,minimum width=12mm]
\tikzstyle{very large dmap}=[draw,doubled,shape=NEbox,inner sep=2pt,minimum height=6mm,fill=white,minimum width=19.5mm]

\tikzstyle{muxbox}=[draw,shape=rectangle,minimum height=3mm,minimum width=3mm,fill=white]
\tikzstyle{dmuxbox}=[muxbox,doubled]

\tikzstyle{box}=[draw,shape=rectangle,inner sep=2pt,minimum height=6mm,minimum width=6mm,fill=white]
\tikzstyle{dbox}=[draw,doubled,shape=rectangle,inner sep=2pt,minimum height=6mm,minimum width=6mm,fill=white]
\tikzstyle{dmap}=[draw,doubled,shape=NEbox,inner sep=2pt,minimum height=6mm,fill=white]
\tikzstyle{dmapdag}=[draw,doubled,shape=SEbox,inner sep=2pt,minimum height=6mm,fill=white]
\tikzstyle{dmapadj}=[draw,doubled,shape=SEbox,inner sep=2pt,minimum height=6mm,fill=white]
\tikzstyle{dmaptrans}=[draw,doubled,shape=SWbox,inner sep=2pt,minimum height=6mm,fill=white]
\tikzstyle{dmapconj}=[draw,doubled,shape=NWbox,inner sep=2pt,minimum height=6mm,fill=white]

\tikzstyle{ddmap}=[draw,doubled,dashed,shape=NEbox,inner sep=2pt,minimum height=6mm,fill=white]
\tikzstyle{ddmapdag}=[draw,doubled,dashed,shape=SEbox,inner sep=2pt,minimum height=6mm,fill=white]
\tikzstyle{ddmapadj}=[draw,doubled,dashed,shape=SEbox,inner sep=2pt,minimum height=6mm,fill=white]
\tikzstyle{ddmaptrans}=[draw,doubled,dashed,shape=SWbox,inner sep=2pt,minimum height=6mm,fill=white]
\tikzstyle{ddmapconj}=[draw,doubled,dashed,shape=NWbox,inner sep=2pt,minimum height=6mm,fill=white]

\boxshape{sNEbox}{0pt}{3pt}
\boxshape{sSEbox}{0pt}{-3pt}
\boxshape{sNWbox}{3pt}{0pt}
\boxshape{sSWbox}{-3pt}{0pt}
\tikzstyle{smap}=[draw,shape=sNEbox,fill=white]
\tikzstyle{smapdag}=[draw,shape=sSEbox,fill=white]
\tikzstyle{smapadj}=[draw,shape=sSEbox,fill=white]
\tikzstyle{smaptrans}=[draw,shape=sSWbox,fill=white]
\tikzstyle{smapconj}=[draw,shape=sNWbox,fill=white]

\tikzstyle{dsmap}=[draw,dashed,shape=sNEbox,fill=white]
\tikzstyle{dsmapdag}=[draw,dashed,shape=sSEbox,fill=white]
\tikzstyle{dsmaptrans}=[draw,dashed,shape=sSWbox,fill=white]
\tikzstyle{dsmapconj}=[draw,dashed,shape=sNWbox,fill=white]

\boxshape{mNEbox}{0pt}{10pt}
\boxshape{mSEbox}{0pt}{-10pt}
\boxshape{mNWbox}{10pt}{0pt}
\boxshape{mSWbox}{-10pt}{0pt}
\tikzstyle{mmap}=[draw,shape=mNEbox]
\tikzstyle{mmapdag}=[draw,shape=mSEbox]
\tikzstyle{mmaptrans}=[draw,shape=mSWbox]
\tikzstyle{mmapconj}=[draw,shape=mNWbox]

\tikzstyle{mmapgray}=[draw,fill=gray!40!white,shape=mNEbox]
\tikzstyle{smapgray}=[draw,fill=gray!40!white,shape=sNEbox]

\makeatletter
\pgfdeclareshape{cornerpoint}{
\inheritsavedanchors[from=rectangle] % this is nearly a rectangle
\inheritanchorborder[from=rectangle]
\inheritanchor[from=rectangle]{center}
\inheritanchor[from=rectangle]{north}
\inheritanchor[from=rectangle]{south}
\inheritanchor[from=rectangle]{west}
\inheritanchor[from=rectangle]{east}
\backgroundpath{% this is new
\southwest \pgf@xa=\pgf@x \pgf@ya=\pgf@y
\northeast \pgf@xb=\pgf@x \pgf@yb=\pgf@y

\pgfmathsetmacro{\pgf@shorten@left}{\pgfkeysvalueof{/tikz/shorten left}}
\pgfmathsetmacro{\pgf@shorten@right}{\pgfkeysvalueof{/tikz/shorten right}}

\pgfpathmoveto{\pgfpoint{0.5 * (\pgf@xa + \pgf@xb)}{\pgf@ya - 5pt}}
\pgfpathlineto{\pgfpoint{\pgf@xa - 8pt + \pgf@shorten@left}{\pgf@yb - 1.5 * \pgf@shorten@left}}
\pgfpathlineto{\pgfpoint{\pgf@xa - 8pt + \pgf@shorten@left}{\pgf@yb}}
\pgfpathlineto{\pgfpoint{\pgf@xb + 8pt - \pgf@shorten@right}{\pgf@yb}}
\pgfpathlineto{\pgfpoint{\pgf@xb + 8pt - \pgf@shorten@right}{\pgf@yb - 1.5 * \pgf@shorten@right}}
\pgfpathclose
}
}

\pgfdeclareshape{cornercopoint}{
\inheritsavedanchors[from=rectangle] % this is nearly a rectangle
\inheritanchorborder[from=rectangle]
\inheritanchor[from=rectangle]{center}
\inheritanchor[from=rectangle]{north}
\inheritanchor[from=rectangle]{south}
\inheritanchor[from=rectangle]{west}
\inheritanchor[from=rectangle]{east}
\backgroundpath{% this is new
\southwest \pgf@xa=\pgf@x \pgf@ya=\pgf@y
\northeast \pgf@xb=\pgf@x \pgf@yb=\pgf@y

\pgfmathsetmacro{\pgf@shorten@left}{\pgfkeysvalueof{/tikz/shorten left}}
\pgfmathsetmacro{\pgf@shorten@right}{\pgfkeysvalueof{/tikz/shorten right}}

\pgfpathmoveto{\pgfpoint{0.5 * (\pgf@xa + \pgf@xb)}{\pgf@yb + 5pt}}
\pgfpathlineto{\pgfpoint{\pgf@xa - 8pt + \pgf@shorten@left}{\pgf@ya + 1.5 * \pgf@shorten@left}}
\pgfpathlineto{\pgfpoint{\pgf@xa - 8pt + \pgf@shorten@left}{\pgf@ya}}
\pgfpathlineto{\pgfpoint{\pgf@xb + 8pt - \pgf@shorten@right}{\pgf@ya}}
\pgfpathlineto{\pgfpoint{\pgf@xb + 8pt - \pgf@shorten@right}{\pgf@ya + 1.5 * \pgf@shorten@right}}
\pgfpathclose
}
}

\pgfdeclareshape{langpoint}{
\inheritsavedanchors[from=rectangle] % this is nearly a rectangle
\inheritanchorborder[from=rectangle]
\inheritanchor[from=rectangle]{center}
\inheritanchor[from=rectangle]{north}
\inheritanchor[from=rectangle]{south}
\inheritanchor[from=rectangle]{west}
\inheritanchor[from=rectangle]{east}
\backgroundpath{% this is new
\southwest \pgf@xa=\pgf@x \pgf@ya=\pgf@y
\northeast \pgf@xb=\pgf@x \pgf@yb=\pgf@y

\pgfmathsetmacro{\pgf@shorten@left}{\pgfkeysvalueof{/tikz/shorten left}}
\pgfmathsetmacro{\pgf@shorten@right}{\pgfkeysvalueof{/tikz/shorten right}}

\pgfpathmoveto{\pgfpoint{0.5 * (\pgf@xa + \pgf@xb)}{\pgf@ya - 2pt}}
\pgfpathlineto{\pgfpoint{\pgf@xa - 8pt}{\pgf@yb - 3 * \pgf@shorten@left + 5pt}} 
\pgfpathlineto{\pgfpoint{\pgf@xa - 8pt}{\pgf@yb -1pt}}
\pgfpathlineto{\pgfpoint{\pgf@xb + 8pt}{\pgf@yb -1pt}}
\pgfpathlineto{\pgfpoint{\pgf@xb + 8pt}{\pgf@yb - 3 * \pgf@shorten@left + 5pt}}%NEW POINT
\pgfpathclose
}
}

\pgfdeclareshape{langcopoint}{
\inheritsavedanchors[from=rectangle] % this is nearly a rectangle
\inheritanchorborder[from=rectangle]
\inheritanchor[from=rectangle]{center}
\inheritanchor[from=rectangle]{north}
\inheritanchor[from=rectangle]{south}
\inheritanchor[from=rectangle]{west}
\inheritanchor[from=rectangle]{east}
\backgroundpath{% this is new
\southwest \pgf@xa=\pgf@x \pgf@ya=\pgf@y
\northeast \pgf@xb=\pgf@x \pgf@yb=\pgf@y

\pgfmathsetmacro{\pgf@shorten@left}{\pgfkeysvalueof{/tikz/shorten left}}
\pgfmathsetmacro{\pgf@shorten@right}{\pgfkeysvalueof{/tikz/shorten right}}

\pgfpathmoveto{\pgfpoint{0.5 * (\pgf@xa + \pgf@xb)}{\pgf@yb +0pt}}
\pgfpathlineto{\pgfpoint{\pgf@xa - 8pt}{\pgf@ya + 3 * \pgf@shorten@left - 5pt}} 
\pgfpathlineto{\pgfpoint{\pgf@xa - 8pt}{\pgf@ya + 1pt}}
\pgfpathlineto{\pgfpoint{\pgf@xb + 8pt}{\pgf@ya + 1pt}}
\pgfpathlineto{\pgfpoint{\pgf@xb + 8pt}{\pgf@ya + 3 * \pgf@shorten@left - 5pt}}%NEW POINT
\pgfpathclose
}
}

\pgfdeclareshape{langrect}{
\inheritsavedanchors[from=rectangle] % this is nearly a rectangle
\inheritanchorborder[from=rectangle]
\inheritanchor[from=rectangle]{center}
\inheritanchor[from=rectangle]{north}
\inheritanchor[from=rectangle]{south}
\inheritanchor[from=rectangle]{west}
\inheritanchor[from=rectangle]{east}
\backgroundpath{% this is new
\southwest \pgf@xa=\pgf@x \pgf@ya=\pgf@y
\northeast \pgf@xb=\pgf@x \pgf@yb=\pgf@y

\pgfmathsetmacro{\pgf@shorten@left}{\pgfkeysvalueof{/tikz/shorten left}}
\pgfmathsetmacro{\pgf@shorten@right}{\pgfkeysvalueof{/tikz/shorten right}}

\pgfpathmoveto{\pgfpoint{\pgf@xa - 8pt}{\pgf@ya + 3 * \pgf@shorten@left - 5pt}} 
\pgfpathlineto{\pgfpoint{\pgf@xa - 8pt}{\pgf@ya + 1pt}}
\pgfpathlineto{\pgfpoint{\pgf@xb + 8pt}{\pgf@ya + 1pt}}
\pgfpathlineto{\pgfpoint{\pgf@xb + 8pt}{\pgf@ya + 3 * \pgf@shorten@left - 5pt}}%NEW POINT
\pgfpathclose
}
}
\makeatother

\pgfkeyssetvalue{/tikz/shorten left}{0pt}
\pgfkeyssetvalue{/tikz/shorten right}{0pt}

\tikzstyle{kpoint common}=[draw,fill=white,inner sep=1pt,minimum height=4mm]

\tikzstyle{langstate}=[shape=langcopoint,shorten left=5pt,kpoint common,font=\footnotesize]
\tikzstyle{langeffect}=[shape=langpoint,shorten left=5pt,kpoint common,font=\footnotesize]
\tikzstyle{langstatedash}=[shape=langcopoint,dashed, shorten left=5pt,kpoint common,font=\footnotesize]
\tikzstyle{langeffectdash}=[shape=langpoint,dashed, shorten left=5pt,kpoint common,font=\footnotesize]
\tikzstyle{langbox}=[shape=langrect,shorten left=5pt,kpoint common,font=\footnotesize] 
\tikzstyle{kpoint}=[shape=cornerpoint,shorten left=5pt,kpoint common]
\tikzstyle{kpoint adjoint}=[shape=cornercopoint,shorten left=5pt,kpoint common]

\tikzstyle{kpoint conjugate}=[shape=cornerpoint,shorten right=5pt,kpoint common]
\tikzstyle{kpoint transpose}=[shape=cornercopoint,shorten right=5pt,kpoint common]
\tikzstyle{kpoint symm}=[shape=cornerpoint,shorten left=5pt,shorten right=5pt,kpoint common]

\tikzstyle{black kpoint}=[shape=cornerpoint,shorten left=5pt,kpoint common,fill=black,font=\color{white}]
\tikzstyle{black kpoint adjoint}=[shape=cornercopoint,shorten left=5pt,kpoint common,fill=black,font=\color{white}]
\tikzstyle{black kpointadj}=[shape=cornercopoint,shorten left=5pt,kpoint common,fill=black,font=\color{white}]

\tikzstyle{black dkpoint}=[shape=cornerpoint,shorten left=5pt,kpoint common,fill=black, doubled,font=\color{white}]
\tikzstyle{black dkpoint adjoint}=[shape=cornercopoint,shorten left=5pt,kpoint common,fill=black, doubled,font=\color{white}]
\tikzstyle{black dkpointadj}=[shape=cornercopoint,shorten left=5pt,kpoint common,fill=black, doubled,font=\color{white}]

\tikzstyle{kpointdag}=[kpoint adjoint]
\tikzstyle{kpointadj}=[kpoint adjoint]
\tikzstyle{kpointconj}=[kpoint conjugate]
\tikzstyle{kpointtrans}=[kpoint transpose]

\tikzstyle{big kpoint}=[kpoint, minimum width=1.2 cm, minimum height=8mm, inner sep=4pt, text depth=3mm]

\tikzstyle{wide kpoint}=[kpoint, minimum width=1 cm, inner sep=2pt]%, text depth=-0.7 mm]
\tikzstyle{wide kpointdag}=[kpointdag, minimum width=1 cm, inner sep=2pt]%, text depth=0.7 mm]
\tikzstyle{wide kpointconj}=[kpointconj, minimum width=1 cm, inner sep=2pt]%, text depth=-0.7 mm]
\tikzstyle{wide kpointtrans}=[kpointtrans, minimum width=1 cm, inner sep=2pt]%, text depth=0.7 mm]

\tikzstyle{gray kpoint}=[kpoint,fill=gray!50!white]
\tikzstyle{gray kpointdag}=[kpointdag,fill=gray!50!white]
\tikzstyle{gray kpointadj}=[kpointadj,fill=gray!50!white]
\tikzstyle{gray kpointconj}=[kpointconj,fill=gray!50!white]
\tikzstyle{gray kpointtrans}=[kpointtrans,fill=gray!50!white]

\tikzstyle{gray dkpoint}=[kpoint,fill=gray!50!white,doubled]
\tikzstyle{gray dkpointdag}=[kpointdag,fill=gray!50!white,doubled]
\tikzstyle{gray dkpointadj}=[kpointadj,fill=gray!50!white,doubled]
\tikzstyle{gray dkpointconj}=[kpointconj,fill=gray!50!white,doubled]
\tikzstyle{gray dkpointtrans}=[kpointtrans,fill=gray!50!white,doubled]

\tikzstyle{white label}=[draw,fill=white,rectangle,inner sep=0.7 mm]
\tikzstyle{gray label}=[draw,fill=gray!50!white,rectangle,inner sep=0.7 mm]
\tikzstyle{black label}=[draw,fill=black,rectangle,inner sep=0.7 mm]

\tikzstyle{dkpoint}=[kpoint,doubled]
\tikzstyle{wide dkpoint}=[wide kpoint,doubled]
\tikzstyle{dkpointdag}=[kpoint adjoint,doubled]
\tikzstyle{wide dkpointdag}=[wide kpointdag,doubled]
\tikzstyle{dkcopoint}=[kpoint adjoint,doubled]
\tikzstyle{dkpointadj}=[kpoint adjoint,doubled]
\tikzstyle{dkpointconj}=[kpoint conjugate,doubled]
\tikzstyle{dkpointtrans}=[kpoint transpose,doubled]

\tikzstyle{kscalar}=[kpoint common, shape=EBox, inner xsep=-1pt, inner ysep=3pt,font=\small]
\tikzstyle{kscalarconj}=[kpoint common, shape=WBox, inner xsep=-1pt, inner ysep=3pt,font=\small]

 \tikzstyle{upground}=[circuit ee IEC,ground,rotate=90,scale=2.5]
 \tikzstyle{downground}=[circuit ee IEC,ground,rotate=-90,scale=2.5]
 \tikzstyle{bigground}=[regular polygon,regular polygon sides=3,draw=gray,scale=0.50,inner sep=-0.5pt,minimum width=10mm,fill=gray]
\tikzstyle{arrs}=[-latex,font=\small,auto]
\tikzstyle{arrow plain}=[arrs]
\tikzstyle{arrow dashed}=[dashed,arrs]
\tikzstyle{arrow bold}=[very thick,arrs]
\tikzstyle{arrow hide}=[draw=white!0,-]
\tikzstyle{arrow reverse}=[latex-]
\tikzstyle{cdnode}=[]

\definecolor{hexcolor0xa9a9a9}{rgb}{0.663,0.663,0.663}
\tikzstyle{GrayLine}=[dashed,draw=hexcolor0xa9a9a9]
\tikzstyle{gray}=[dashed,draw=hexcolor0xa9a9a9]

\theoremstyle{definition}

\newtheorem*{theorem*}{Theorem}

\newtheorem{example*}[theorem]{Example*}
\newtheorem{examples*}[theorem]{Examples*}

\newtheorem{remark*}[theorem]{Remark*}

\def\bR{\begin{color}{red}}
\def\bB{\begin{color}{blue}}
\def\bM{\begin{color}{black}}
\def\bC{\begin{color}{cyan}}
\def\bW{\begin{color}{white}}
\def\bMl{\begin{color}{black}}
\def\bG{\begin{color}{green}}
\def\bY{\begin{color}{yellow}}
\def\e{\end{color}\xspace}
\newcommand{\bit}{\begin{itemize}}
\newcommand{\eit}{\end{itemize}\par\noindent}
\newcommand{\ben}{\begin{enumerate}}
\newcommand{\een}{\end{enumerate}\par\noindent}
\newcommand{\beq}{\begin{equation}}
\newcommand{\eeq}{\end{equation}\par\noindent}
\newcommand{\beqa}{\begin{eqnarray*}}
\newcommand{\eeqa}{\end{eqnarray*}\par\noindent}
\newcommand{\beqn}{\begin{eqnarray}}
\newcommand{\eeqn}{\end{eqnarray}\par\noindent}

\title{\bf How to make qubits speak
}
\author{Bob Coecke, Giovanni de Felice, Konstantinos Meichanetzidis, Alexis Toumi\\ \ \\
Cambridge Quantum Computing Ltd.  \\
Oxford University, Department of Computer Science\\ \ \\
{\bM Invited contribution to:\e}\\ {\bM\it``Quantum Computing in the Arts and Humanities"\e}\\
}
\begin{document}
\maketitle

\begin{abstract}
This is a story about making quantum computers speak, and doing so in a quantum-native, compositional and meaning-aware manner.  Recently we did question-answering with an actual quantum computer.  We explain what we did, stress that this was all done in terms of pictures, and provide many pointers to the related literature.  In fact, besides natural language, many other things can be implemented in a quantum-native, compositional and meaning-aware manner, and we provide the reader with some indications of that broader pictorial landscape, \bM including our account on the notion of compositionality.  We also provide some guidance  for the actual execution, so that the reader can give it a go as well. \e
\end{abstract}

\section{Networks of words}

What is a word?

In particular, what is the meaning of a word?  This could be what you find in a dictionary. Then again, before there were dictionaries people also used words, and so did people that could not read.  In fact, most words we know we didn't look up in a dictionary.
We probably learned them by hearing them used a lot.   This idea of meaning is closer to how machines learn meanings of words today: they learn them from the context in which they are used, an idea coined by Wittgenstein as ``meaning is use'', and expressed by Firth's derived dictum \cite{Firth}:
\begin{center}
``You shall know a word by the company it keeps".
\end{center}

\noindent
What is a sentence?

In particular, what is the meaning of a sentence? A sentence is not just a ``bag of words",\footnote{Here, `bag' is a widely used technical term, referring to the fact that in many `natural language processing' applications grammatical structure has been entirely ignored for many years \cite{harris1954distributional}.} but rather, a kind of network in which words interact in a particular fashion. In fact, in a very particular fashion, given that when we hear a sentence that we never heard before, provided we do understand the words that occur in it, then we surely also understand that sentence.  That's exactly why we don't have dictionaries for sentences---besides the fact a `sentence dictionary' would take up an entire library, if not more.  There is an important academic question here:
\bit
\item How do we deduce the meaning of a sentence given the meanings of its words?
\eit
We can also ask the converse question:
\bit
\item Can we infer the meanings of words in sentences, from the meanings of sentences?
\eit
Both turn out to also be essential practical questions.

Some 10 years ago, in  \cite{CSC}, BC Mehrnoosh Sadrzadeh and Steve Clark started to draw  networks in order to address the first of these questions. These networks look like this:
\beq
\begin{tikzpicture}[scale=0.80]
	\begin{pgfonlayer}{nodelayer}
		\node [style=none] (0) at (-9.25, -0.25) {};
		\node [style=none] (1) at (10.75, -0.25) {};
		\node [style=none] (2) at (8, -0.25) {};
		\node [style=langstate] (3) at (-3.5, 0.25) {\!\!\!{\tt flowers}\!\!\!};
		\node [style=langstate] (4) at (-9.25, 0.25) {\!\!\!{\tt likes}\!\!\!};
		\node [style=langstate] (5) at (4.25, 0.25) {\!\!\!{\tt Bob}\!\!\!};
		\node [style=langstate] (6) at (7.25, 0.25) {\!\!\!{\tt gives}\!\!\!};
		\node [style=langstate] (7) at (10.75, 0.25) {\!\!\!{\tt Claire}\!\!\!};
		\node [style=none] (8) at (6.5, -0.25) {};
		\node [style=none] (9) at (4.25, -0.25) {};
		\node [style=none] (10) at (-3.5, -0.25) {};
		\node [style=none] (11) at (-0.75, -0.25) {};
		\node [style=none] (12) at (0.25, -0.25) {};
		\node [style=none] (13) at (-8.5, -0.25) {};
		\node [style=none] (14) at (1.25, -0.25) {};
		\node [style=none] (15) at (7.5, -0.25) {};
		\node [style=none] (16) at (2.25, -0.25) {};
		\node [style=none] (17) at (7, -0.25) {};
		\node [style=none] (18) at (0.25, -0.25) {};
		\node [style=none] (19) at (-0.75, -0.25) {};
		\node [style=none] (20) at (-0.75, -0.25) {};
		\node [style=none] (21) at (0.25, -0.25) {};
		\node [style=none] (22) at (0.25, -0.25) {};
		\node [style=none] (23) at (0.25, -0.25) {};
		\node [style=none] (24) at (2.5, -0.25) {};
		\node [style=none] (25) at (1.25, -0.25) {};
		\node [style=none] (26) at (1.25, -0.25) {};
		\node [style=none] (27) at (-1.25, -0.25) {};
		\node [style=none] (28) at (2.75, 0.75) {};
		\node [style=none] (29) at (1.25, -0.25) {};
		\node [style=none] (30) at (1.25, -0.25) {};
		\node [style=none] (31) at (-0.75, -0.25) {};
		\node [style=none] (32) at (2.25, -0.25) {};
		\node [style=none] (33) at (2.75, -0.25) {};
		\node [style=none] (34) at (-1.25, 0.75) {};
		\node [style=none] (35) at (0.25, -0.25) {};
		\node [style=none] (36) at (0.25, -0.25) {};
		\node [style=none] (37) at (0.25, -0.25) {};
		\node [style=white dot] (38) at (0.25, 0.75) {};
		\node [style=none] (39) at (1.25, -0.25) {};
		\node [style=none] (40) at (0.25, -0.25) {};
		\node [style=none] (41) at (0.75, 1.25) {};
		\node [style=none] (42) at (-0.75, -0.25) {};
		\node [style=none] (43) at (2.25, -0.25) {};
		\node [style=none] (44) at (2.25, -0.25) {};
		\node [style=none] (45) at (0.75, 1.75) {\!\!\!{\footnotesize\tt that}\!\!\!};
		\node [style=none] (46) at (-12.5, -0.25) {};
		\node [style=none] (47) at (-10, -0.25) {};
		\node [style=none] (48) at (-9.25, -0.25) {};
		\node [style=none] (49) at (-6.5, 0.25) {\footnotesize\tt (the)};
		\node [style=langstate] (50) at (-12.5, 0.25) {\!\!\!{\tt Alice}\!\!\!};
		\node [style=none] (51) at (-10, -0.25) {};
		\node [style=none] (52) at (-9.25, -0.25) {};
		\node [style=none] (53) at (-9.25, -0.25) {};
		\node [style=none] (54) at (-9.25, -1.75) {};
		\node [style=none] (55) at (6.5, 0.25) {};
		\node [style=none] (56) at (-10, 0.25) {};
		\node [style=none] (57) at (-8.5, 0.25) {};
		\node [style=none] (58) at (7.5, 0.25) {};
		\node [style=none] (59) at (7, 0.25) {};
		\node [style=none] (60) at (8, 0.25) {};
		\node [style=white ddot] (61) at (2.25, 0.25) {};
	\end{pgfonlayer}
	\begin{pgfonlayer}{edgelayer}
		\draw [style=swap, bend right=90, looseness=1.00] (2.center) to (1.center);
		\draw [style=swap, bend right=90, looseness=1.00] (9.center) to (8.center);
		\draw [style=swap, bend right=90, looseness=0.75] (10.center) to (11.center);
		\draw [style=swap, bend right=90, looseness=0.50] (13.center) to (12.center);
		\draw [style=swap, bend right=90, looseness=0.75] (14.center) to (15.center);
		\draw [style=swap, in=180, out=90, looseness=1.00] (42.center) to (38);
		\draw [style=swap] (38) to (36.center);
		\draw [style=swap, in=90, out=0, looseness=1.00] (38) to (30.center);
		\draw [style=swap, dashed] (34.center) to (27.center);
		\draw [style=swap, dashed] (27.center) to (33.center);
		\draw [style=swap, dashed] (33.center) to (28.center);
		\draw [style=swap, dashed] (41.center) to (34.center);
		\draw [style=swap, dashed] (41.center) to (28.center);
		\draw [style=swap, bend right=90, looseness=1.00] (46.center) to (47.center);
		\draw [style=boldedge] (53.center) to (54.center);
		\draw [style=swap] (7) to (1.center);
		\draw [style=swap] (5) to (9.center);
		\draw [style=swap] (3) to (10.center);
		\draw [style=swap] (4) to (52.center);
		\draw [style=swap] (50) to (46.center);
		\draw [style=swap] (56.center) to (51.center);
		\draw [style=swap] (57.center) to (13.center);
		\draw [style=swap] (55.center) to (8.center);
		\draw [style=swap] (58.center) to (15.center);
		\draw [style=swap] (60.center) to (2.center);
		\draw [style=boldedge] (61) to (44.center);
		\draw [style=boldedge, bend right=90, looseness=0.75] (16.center) to (17.center);
		\draw [style=boldedge] (59.center) to (17.center);
		\draw [style=boldedge] (4) to (52.center);
	\end{pgfonlayer}
\end{tikzpicture}
\eeq
The usual technical term for these networks is ``string diagram'' (see e.g.~\cite{BaezLNP}).  These string diagrams provided a straight-forward manner for how  meanings of  words  combine in order to produce the meaning of a sentence,
%as opposed to treating the sentence as a structureless ``bag" containing the meanings of individual words, and even
even resulting in a cover-heading feature in New Scientist.\footnote{https://www.newscientist.com/article/mg20827903-200-quantum-links-let-computers-understand-language/}

In order to better see how these string diagrams do so, let's consider a  simpler example:
\beq\label{eq:tv2wg}
\begin{tikzpicture}[scale=0.80]
	\begin{pgfonlayer}{nodelayer}
		\node [style=none] (0) at (-4, 0.5) {};
		\node [style=langstate] (1) at (-4, 0.75) {\!\!\!{\tt Alice}\!\!\!};
		\node [style=langstate] (2) at (4, 0.75) {\!\!\!{\tt Bob}\!\!\!};
		\node [style=none] (3) at (4, 0.5) {};
		\node [style=langstate] (4) at (0, 0.75) {\ {\tt hates}\ };
		\node [style=none] (5) at (0, -1) {};
		\node [style=none] (6) at (1.5, 0.5) {};
		\node [style=none] (7) at (0, 0.5) {};
		\node [style=none] (8) at (-1.5, 0.5) {};
		\node [style=none] (9) at (4, 0.5) {};
		\node [style=none] (10) at (-1.5, 0.5) {};
		\node [style=none] (11) at (1.5, 0.5) {};
		\node [style=none] (12) at (1.5, 0.5) {};
		\node [style=none] (13) at (-1.5, 0.5) {};
		\node [style=none] (14) at (-4, 0.5) {};
		\node [style=none] (15) at (-1.5, 0.5) {};
	\end{pgfonlayer}
	\begin{pgfonlayer}{edgelayer}
		\draw [style=boldedge] (7.center) to (5.center);
		\draw [style=swap, bend right=90, looseness=1.50] (11.center) to (9.center);
		\draw [style=swap, bend left=90, looseness=1.50] (10.center) to (14.center);
	\end{pgfonlayer}
\end{tikzpicture}
\eeq
The idea here is that the boxes represent the meanings of words and that the wires are channels through which these meanings can be transmitted. So, in the example above, the subject {\tt Alice} and the object {\tt Bob} are both sent to the verb {\tt hates}, and together they then make up the meaning of the sentence. This idea scales to much bigger sentences like the one depicted above, and even to large text made up of multiple sentences \cite{CoeckeText}.

This flow of words in sentences can be traced back to work originally started in the 1930s by Adjukiewicz \cite{Adjukiewicz35} and later by Bar-Hillel
\cite{Bar-Hillel}, Chomsky \cite{Chomsky, ChomskyBook1} and Lambek \cite{Lambek0, LambekBook}, among others.  What they did was unifying grammatical structures across different languages as a single mathematical structure.  An overview of these developments from the perspective of string diagrams is in \cite{Gospel}.
%In particular, the language dia of a sentence's meaning-flow is constructed according to a compositional mathematical model of meaning (a.k.a.~semantics) \cite{Gospel}.

\section{Language is quantum-native}

One particularly interesting aspect of this graphical framework for linguistics was that the string diagrams were inherited from previous work that provided quantum theory with a network-like language \cite{Kindergarten}. This work was accumulated eventually in a 900-page book
%written by BC and Aleks Kissinger
\cite{CKbook}.

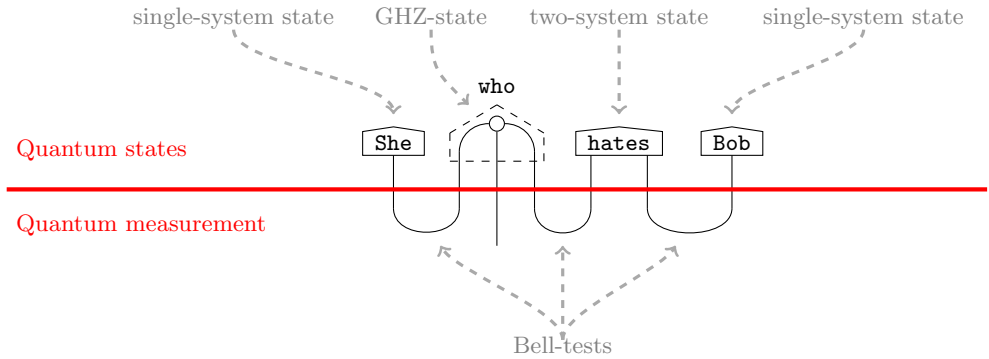
\begin{figure}
\centering
\begin{tikzpicture}[scale=1]
	\begin{pgfonlayer}{nodelayer}
		\node [style=none] (0) at (0.5, -1.75) {};
		\node [style=langstate] (1) at (-4.75, 0) {\!\!\!{\tt She}\!\!\!};
		\node [style=none] (2) at (4.25, -1.75) {};
		\node [style=none] (3) at (-1, -0.25) {};
		\node [style=none] (4) at (0.5, -0.25) {};
		\node [style=langstate] (5) at (1.25, 0) {\!\!\!{\tt hates}\!\!\!};
		\node [style=none] (6) at (-2, -2.75) {};
		\node [style=none] (7) at (2, -1.75) {};
		\node [style=none] (8) at (-1, -1.75) {};
		\node [style=none] (9) at (-2, 1.5) {\footnotesize\tt who};
		\node [style=none] (10) at (0.5, -1.75) {};
		\node [style=none] (11) at (-3, -0.25) {};
		\node [style=none] (12) at (2, -0.25) {};
		\node [style=none] (13) at (-3, -1.75) {};
		\node [style=none] (14) at (-4.75, -1.75) {};
		\node [style=langstate] (15) at (4.25, 0) {\!\!\!{\tt Bob}\!\!\!};
		\node [style=white dot] (16) at (-2, 0.5) {};
		\node [style=none] (17) at (-3.25, 0.25) {};
		\node [style=none] (18) at (-3.25, -0.5) {};
		\node [style=none] (19) at (-2, 1) {};
		\node [style=none] (20) at (-0.75, -0.5) {};
		\node [style=none] (21) at (-0.75, 0.25) {};
		\node [style=none] (22) at (4.25, -0.25) {};
		\node [style=none] (23) at (-4.75, -0.25) {};
		\node [style=none] (24) at (1.25, 3) {};
		\node [style=label] (25) at (1.25, 3.25) {\color{gray}two-system state};
		\node [style=none] (26) at (1.25, 0.75) {};
		\node [style=none] (27) at (4.25, 0.75) {};
		\node [style=label] (28) at (7.75, 3.25) {\color{gray}single-system state};
		\node [style=none] (29) at (7.75, 3) {};
		\node [style=label] (30) at (-9, 3.25) {\color{gray}single-system state};
		\node [style=none] (31) at (-9, 3) {};
		\node [style=none] (32) at (-4.75, 0.75) {};
		\node [style=none] (33) at (-0.25, -2.75) {};
		\node [style=none] (34) at (-0.25, -5.25) {};
		\node [style=label] (35) at (-0.25, -5.5) {\color{gray}Bell-tests};
		\node [style=none] (36) at (-0.25, -5.25) {};
		\node [style=none] (37) at (2.75, -2.75) {};
		\node [style=none] (38) at (-0.25, -5.25) {};
		\node [style=none] (39) at (-3.5, -2.75) {};
		\node [style=label] (40) at (-3.75, 3.25) {\color{gray}GHZ-state};
		\node [style=none] (41) at (-3.75, 3) {};
		\node [style=none] (42) at (-2.75, 1) {};
		\node [style=none] (43) at (11, -1.25) {};
		\node [style=none] (44) at (-15, -1.25) {};
		\node [style=right label, red] (45) at (-14.75, -0.25) {Quantum states};
		\node [style=right label, red] (46) at (-14.75, -2.25) {Quantum measurement};
	\end{pgfonlayer}
	\begin{pgfonlayer}{edgelayer}
		\draw [style=swap] (7.center) to (12.center);
		\draw [style=swap, bend right=90, looseness=1.00] (7.center) to (2.center);
		\draw [style=swap] (10.center) to (4.center);
		\draw [style=swap] (13.center) to (11.center);
		\draw [style=swap] (8.center) to (3.center);
		\draw [style=swap, bend right=90, looseness=1.50] (8.center) to (0.center);
		\draw [style=swap, bend right=90, looseness=1.25] (14.center) to (13.center);
		\draw [style=swap, in=0, out=90, looseness=1.00] (3.center) to (16);
		\draw [style=swap, in=-90, out=90, looseness=1.00] (6.center) to (16);
		\draw [style=swap, in=180, out=90, looseness=1.00] (11.center) to (16);
		\draw [style=swap, dashed] (17.center) to (18.center);
		\draw [style=swap, dashed] (18.center) to (20.center);
		\draw [style=swap, dashed] (20.center) to (21.center);
		\draw [style=swap, dashed] (19.center) to (17.center);
		\draw [style=swap, dashed] (19.center) to (21.center);
		\draw [style=swap] (23.center) to (14.center);
		\draw [style=swap] (22.center) to (2.center);
		\draw [style=pointer edge, in=90, out=-90, looseness=0.50] (24.center) to (26.center);
		\draw [style=pointer edge, in=90, out=-90, looseness=0.50] (29.center) to (27.center);
		\draw [style=pointer edge, in=90, out=-90, looseness=0.50] (31.center) to (32.center);
		\draw [style=pointer edge, in=-90, out=90, looseness=0.50] (34.center) to (33.center);
		\draw [style=pointer edge, in=-120, out=90, looseness=0.50] (36.center) to (37.center);
		\draw [style=pointer edge, in=-60, out=90, looseness=0.50] (38.center) to (39.center);
		\draw [style=pointer edge, in=135, out=-90, looseness=1.25] (41.center) to (42.center);
		\draw [style=boldedge, red] (44.center) to (43.center);
	\end{pgfonlayer}
\end{tikzpicture}
\bM\caption{Illustration of how word meanings can be interpreted as quantum states, and grammatical structure in terms of quantum measurement.}\label{Qinterfig}\e
\end{figure}

In summary, a direct correspondence was established between, on the one hand, the meanings of words and quantum states, and on the other hand, grammatical structures and quantum measurements. \bM This is illustrated in Figure \ref{Qinterfig}. \e
Obviously this led to the question: Can one make quantum computers handle natural language? This was first proposed in a paper by Will Zeng and BC in 2016 \cite{WillC},  creating a new paradigm for natural language processing (NLP) in a quantum computing context.

The idea of making quantum computers process natural language is just not only incredibly cool but also a very natural thing to do for the reasons indicated above. More researchers started to take an interest, and at some point Intel supported a basic attempt to develop some of the ideas contained in \cite{WillC} on their quantum simulator \cite{INTEL, o2020hybrid}.

However, there were some significant challenges to the proposal. Most importantly, there weren't any sufficiently capable quantum computers that could implement the NLP tasks proposed. Additionally, an assumption was made that one could encode word meanings on the quantum computer using quantum random access memory (QRAM) \cite{giovannetti2008quantum}, which to this day, and despite theoretical progress and experimental proposals, remains a distant possibility.

\section{Our new attempt}

%The first conference on Quantum Natural Language Processing, or ?QNLP?as we?ve called it, took place in Oxford in December 2019,�� where we presented a simulation of our experiment�? (all talks can be found online��).

In the past year, we have been examining ways to use existing NISQ  (= Noisy Intermediate Scale Quantum) devices, in the first instance one of IBM's quantum devices, for NLP.

The string diagrams as depicted above can't be interpreted directly by IBM's machine, which instead, needs something in the form of a `quantum circuit'. Natural language when we map it to a `quantum circuit skeleton' now looks like this \cite{QNLP-foundations}:
\beq\label{tv17more}
\begin{tikzpicture}[scale=0.80]
	\begin{pgfonlayer}{nodelayer}
		\node [style=gray dot] (0) at (1.75, 3) {};
		\node [style=gray dot] (1) at (1.75, -8) {};
		\node [style=gray dot] (2) at (1.75, 7.25) {};
		\node [style=none] (3) at (1.75, -8) {};
		\node [style=none] (4) at (5.25, 4.5) {};
		\node [style=none] (5) at (5.25, 3) {};
		\node [style=none] (6) at (5.25, 6) {};
		\node [style=none] (7) at (-5.25, -7) {};
		\node [style=gray dot] (8) at (1.75, -7) {};
		\node [style=none] (9) at (1.75, 1.5) {};
		\node [style=none] (10) at (1.75, 3) {};
		\node [style=none] (11) at (5.25, 3) {};
		\node [style=white dot] (12) at (-5.25, -7) {}; 
		\node [style=none] (13) at (-5.25, 4.5) {};
		\node [style=none] (14) at (1.75, -7) {};
		\node [style=none] (15) at (-5.25, -8.5) {};
		\node [style=none] (16) at (-5.25, -7) {};
		\node [style=white dot] (17) at (5.25, 3) {};
		\node [style=none] (18) at (5.25, -8.5) {};
		\node [style=none] (19) at (1.75, 1.5) {};
		\node [style=white phase dot] (20) at (1.75, 1.5) {$\alpha$};
		\node [style=none] (21) at (1.75, 0) {};
		\node [style=none] (22) at (1.75, 0) {};
		\node [style=none] (23) at (1.75, 0) {};
		\node [style=none] (24) at (1.75, 0) {};
		\node [style=gray phase dot] (25) at (1.75, 0) {$\beta$};
		\node [style=none] (26) at (1.75, -4) {};
		\node [style=white phase dot] (27) at (1.75, -2.5) {$\alpha'\! +\gamma$};
		\node [style=gray phase dot] (28) at (1.75, -4) {$\beta'$};
		\node [style=none] (29) at (1.75, -4) {};
		\node [style=none] (30) at (1.75, -2.5) {};
		\node [style=none] (31) at (1.75, -4) {};
		\node [style=none] (32) at (1.75, -5.5) {};
		\node [style=none] (33) at (1.75, -5.5) {};
		\node [style=none] (34) at (1.75, -5.5) {};
		\node [style=none] (35) at (1.75, -5.5) {};
		\node [style=none] (36) at (1.75, -2.5) {};
		\node [style=none] (37) at (1.75, -4) {};
		\node [style=white phase dot] (38) at (1.75, -5.5) {$\gamma'$};
		\node [style=white dot] (39) at (1.75, -1.25) {};
		\node [style=none] (40) at (-1.75, -1.25) {};
		\node [style=none] (41) at (1.75, -1.25) {};
		\node [style=none] (42) at (1.75, -1.25) {};
		\node [style=gray dot] (43) at (-1.75, -1.25) {};
		\node [style=gray dot] (44) at (-1.75, -2.25) {};
		\node [style=none] (45) at (-1.75, -2.25) {};
		\node [style=white phase dot] (46) at (-5.25, 4.5) {$\beta_A$};
		\node [style=none] (47) at (-5.25, 6) {};
		\node [style=gray phase dot] (48) at (-5.25, 6) {$\alpha_A$};
		\node [style=none] (49) at (-5.25, 4.5) {};
		\node [style=gray dot] (50) at (-5.25, 7.25) {};
		\node [style=gray dot] (51) at (-1.75, 7.25) {};
		\node [style=gray phase dot] (52) at (-1.75, 6) {$\alpha_{\bf p}$};
		\node [style=none] (53) at (-1.75, 6) {};
		\node [style=none] (54) at (-1.75, 6) {};
		\node [style=none] (55) at (-1.75, 6) {};
		\node [style=none] (56) at (-1.75, 6) {};
		\node [style=none] (57) at (5.25, 6) {};
		\node [style=white phase dot] (58) at (5.25, 4.5) {$\beta_B$};
		\node [style=none] (59) at (5.25, 6) {};
		\node [style=none] (60) at (5.25, 6) {};
		\node [style=none] (61) at (5.25, 6) {};
		\node [style=none] (62) at (5.25, 6) {};
		\node [style=gray dot] (63) at (5.25, 7.25) {};
		\node [style=none] (64) at (5.25, 6) {};
		\node [style=none] (65) at (5.25, 6) {};
		\node [style=none] (66) at (5.25, 6) {};
		\node [style=gray phase dot] (67) at (5.25, 6) {$\alpha_B$};
		\node [style=none] (68) at (5.25, 6) {};
		\node [style=none] (69) at (-2.75, -6.5) {};
		\node [style=none] (70) at (-2.75, 7.75) {};
		\node [style=none] (71) at (3.25, -6.5) {};
		\node [style=none] (72) at (0, 7.75) {};
		\node [style=none] (73) at (0, 2.5) {};
		\node [style=none] (74) at (3.25, 2.5) {};
		\node [style=none] (75) at (-4.25, 7.75) {};
		\node [style=none] (76) at (-6.25, 7.75) {};
		\node [style=none] (77) at (-4.25, 3.5) {};
		\node [style=none] (78) at (-6.25, 3.5) {};
		\node [style=none] (79) at (6.25, 7.75) {};
		\node [style=none] (80) at (4.25, 7.75) {};
		\node [style=none] (81) at (6.25, 3.5) {};
		\node [style=none] (82) at (4.25, 3.5) {};
		\node [style=none] (83) at (3.75, -2) {};
		\node [style=none] (84) at (8.75, -2) {};
		\node [style=right label] (85) at (9, -2) {{\tt hates}};
		\node [style=none] (86) at (8.75, 5.75) {};
		\node [style=right label] (87) at (9, 5.75) {{\tt Bob}};
		\node [style=none] (88) at (6.75, 5.75) {};
		\node [style=none] (89) at (-8.75, 5.75) {};
		\node [style=left label] (90) at (-9, 5.75) {{\tt Alice}};
		\node [style=none] (91) at (-6.75, 5.75) {};
	\end{pgfonlayer}
	\begin{pgfonlayer}{edgelayer} 
		\draw [style=swap] (7.center) to (13.center);
		\draw [style=swap] (16.center) to (15.center);
		\draw [style=swap, in=0, out=180, looseness=1.25] (14.center) to (7.center);
		\draw [style=swap] (8) to (1);
		\draw [style=swap] (5.center) to (4.center);
		\draw [style=swap] (11.center) to (18.center);
		\draw [style=swap, in=180, out=0, looseness=1.25] (10.center) to (5.center);
		\draw [style=swap] (0) to (2);
		\draw [style=swap] (0) to (9.center);
		\draw [style=swap] (20) to (21.center);
		\draw [style=swap] (27) to (26.center);
		\draw [style=swap] (26.center) to (33.center);
		\draw [style=swap, in=180, out=0, looseness=1.25] (40.center) to (42.center);
		\draw [style=swap] (39) to (36.center);
		\draw [style=swap] (32.center) to (14.center);
		\draw [style=swap] (43) to (44);
		\draw [style=swap] (50) to (47.center);
		\draw [style=swap] (63) to (59.center);
		\draw [style=swap] (51) to (52);
		\draw [style=swap] (52) to (43);
		\draw [style=swap] (6.center) to (4.center);
		\draw [style=swap] (47.center) to (46);
		\draw [style=swap] (25) to (39);
		\draw [style=swap, dashed] (69.center) to (70.center);
		\draw [style=swap, dashed] (70.center) to (72.center);
		\draw [style=swap, dashed] (69.center) to (71.center);
		\draw [style=swap, dashed] (72.center) to (73.center);
		\draw [style=swap, dashed] (73.center) to (74.center);
		\draw [style=swap, dashed] (74.center) to (71.center);
		\draw [style=swap, dashed] (76.center) to (75.center);
		\draw [style=swap, dashed] (78.center) to (77.center);
		\draw [style=swap, dashed] (76.center) to (78.center);
		\draw [style=swap, dashed] (75.center) to (77.center);
		\draw [style=swap, dashed] (80.center) to (79.center);
		\draw [style=swap, dashed] (82.center) to (81.center);
		\draw [style=swap, dashed] (80.center) to (82.center);
		\draw [style=swap, dashed] (79.center) to (81.center);
		\draw [style=pointer edge, in=0, out=180, looseness=0.75] (84.center) to (83.center);
		\draw [style=pointer edge, in=0, out=180, looseness=0.75] (86.center) to (88.center);
		\draw [style=pointer edge, in=180, out=0, looseness=0.75] (89.center) to (91.center);
	\end{pgfonlayer}
\end{tikzpicture}
\eeq
In this picture as well as in our experiments we used the `ZX-calculus' \cite{CD2} for drawing quantum circuits, which is again part of the same string diagram language of quantum theory\bM---we give the translation of ZX-calculus circuit components in standard circuit gates in Figure \ref{ZXstandard}\e.
\begin{figure}
\centering
\bM$
\begin{tabular}{|c|c|c|c|c|}
$X$-preparation & $X$-phase gate & $Z$-phase gate & CNOT gate & $X$-measurement\\
\hline
\begin{tikzpicture}[scale=0.80]
	\begin{pgfonlayer}{nodelayer}
		\node [style=gray dot] (0) at (0, 0.5) {};
		\node [style=none] (1) at (0, -0.5) {};
	\end{pgfonlayer}
	\begin{pgfonlayer}{edgelayer}
		\draw [style=swap] (0) to (1.center);
	\end{pgfonlayer}
\end{tikzpicture} & \begin{tikzpicture}[scale=0.80]
	\begin{pgfonlayer}{nodelayer}
		\node [style=none] (0) at (0, 1.5) {};
		\node [style=none] (1) at (0, 0) {};
		\node [style=white phase dot] (2) at (0, 0) {$\alpha$};
		\node [style=none] (3) at (0, -1.5) {};
		\node [style=none] (4) at (0, 1.75) {};
		\node [style=none] (5) at (0, -1.75) {};
	\end{pgfonlayer}
	\begin{pgfonlayer}{edgelayer}
		\draw [style=swap] (2) to (3.center);
		\draw [style=swap] (0.center) to (1.center);
	\end{pgfonlayer}
\end{tikzpicture} & \begin{tikzpicture}[scale=0.80]
	\begin{pgfonlayer}{nodelayer}
		\node [style=none] (0) at (0, 1.5) {};
		\node [style=none] (1) at (0, 0) {};
		\node [style=gray phase dot] (2) at (0, 0) {$\alpha$};
		\node [style=none] (3) at (0, -1.5) {};
	\end{pgfonlayer}
	\begin{pgfonlayer}{edgelayer}
		\draw [style=swap] (2) to (3.center);
		\draw [style=swap] (0.center) to (1.center);
	\end{pgfonlayer}
\end{tikzpicture} & \begin{tikzpicture}[scale=0.80]
	\begin{pgfonlayer}{nodelayer}
		\node [style=none] (0) at (-1, 1) {};
		\node [style=gray dot] (1) at (1.25, 0) {};
		\node [style=none] (2) at (1.25, 1) {};
		\node [style=none] (3) at (-1, -1) {};
		\node [style=none] (4) at (1.25, -1) {};
		\node [style=white dot] (5) at (-1, 0) {};
	\end{pgfonlayer}
	\begin{pgfonlayer}{edgelayer}
		\draw [style=swap] (0.center) to (5); 
		\draw [style=swap] (5) to (3.center);
		\draw [style=swap] (5) to (1);
		\draw [style=swap] (2.center) to (1);
		\draw [style=swap] (1) to (4.center);
	\end{pgfonlayer}
\end{tikzpicture} & \begin{tikzpicture}[scale=0.80]
	\begin{pgfonlayer}{nodelayer}
		\node [style=gray dot] (0) at (0, -0.5) {};
		\node [style=none] (1) at (0, 0.5) {};
	\end{pgfonlayer}
	\begin{pgfonlayer}{edgelayer}
		\draw [style=swap] (0) to (1.center);
	\end{pgfonlayer}
\end{tikzpicture}\\
\hline
 $\begin{pmatrix}
 1\\
 0
 \end{pmatrix}$
 &
$\begin{pmatrix}
 1 & 0\\
 0 & e^{i\alpha}
 \end{pmatrix}$
 &
 $H\circ\begin{pmatrix}
 1 & 0\\
 0 & e^{i\alpha}
 \end{pmatrix}\circ H$
 &
 $\begin{pmatrix}
 1 & 0 & 0 & 0\\
 0 & 1 & 0 & 0\\
 0 & 0 & 0 & 1\\
 0 & 0 & 1 & 0\\
 \end{pmatrix}$
 &
 $\begin{pmatrix}
 1 & 0
 \end{pmatrix}$
 \\
\end{tabular}
$
\caption{Translation of ZX-calculus circuit components in standard circuit gates.  For a short tutorial see \cite{coecke2012tutorial}, or for a more extensive ones see \cite{CKbook, JohnSurvey}.}\label{ZXstandard}\e
\end{figure}
The building blocks of the ZX-language are these `spiders':
\beq
\begin{tikzpicture}[scale=0.80]
	\begin{pgfonlayer}{nodelayer}
		\node [style=none] (0) at (1, -1.25) {};
		\node [style=none] (1) at (-1, -1.25) {};
		\node [style=none] (2) at (-1, 1.25) {};
		\node [style=none] (3) at (1, 1.25) {};
		\node [style=none] (4) at (0, 1) {\ldots};
		\node [style=none] (5) at (0, -1) {\ldots};
		\node [style=white phase dot] (6) at (0, 0) {$\alpha$};
		\node [style=none] (7) at (0, 0) {};
	\end{pgfonlayer}
	\begin{pgfonlayer}{edgelayer}
		\draw [style=swap, in=135, out=-90, looseness=1.00] (2.center) to (6);
		\draw [style=swap, in=45, out=-90, looseness=1.00] (3.center) to (6);
		\draw [style=swap, in=90, out=-45, looseness=1.00] (6) to (0.center);
		\draw [style=swap, in=90, out=-135, looseness=1.00] (6) to (1.center);
	\end{pgfonlayer}
\end{tikzpicture} \qquad\qquad\qquad\qquad\begin{tikzpicture}[scale=0.80]
	\begin{pgfonlayer}{nodelayer}
		\node [style=none] (0) at (1, -1.25) {};
		\node [style=none] (1) at (-1, -1.25) {};
		\node [style=none] (2) at (-1, 1.25) {};
		\node [style=none] (3) at (1, 1.25) {};
		\node [style=none] (4) at (0, 1) {\ldots};
		\node [style=none] (5) at (0, -1) {\ldots};
		\node [style=gray phase dot] (6) at (0, 0) {$\alpha$};
		\node [style=none] (7) at (0, 0) {};
	\end{pgfonlayer}
	\begin{pgfonlayer}{edgelayer}
		\draw [style=swap, in=135, out=-90, looseness=1.00] (2.center) to (6);
		\draw [style=swap, in=45, out=-90, looseness=1.00] (3.center) to (6);
		\draw [style=swap, in=90, out=-45, looseness=1.00] (6) to (0.center);
		\draw [style=swap, in=90, out=-135, looseness=1.00] (6) to (1.center);
	\end{pgfonlayer}
\end{tikzpicture}
\eeq
and rules for computing and reasoning look as follows:
\beq
\begin{tikzpicture}[scale=0.80]
	\begin{pgfonlayer}{nodelayer}
		\node [style=none] (0) at (0, 0) {};
		\node [style=none] (1) at (-2, 0) {};
		\node [style=none] (2) at (-2, 2.5) {};
		\node [style=none] (3) at (0, 2.5) {};
		\node [style=none] (4) at (-1, 2.25) {\ldots};
		\node [style=none] (5) at (-1, 0.25) {\ldots};
		\node [style=white phase  dot] (6) at (-1, 1.25) {$\alpha$};
		\node [style=none] (7) at (-1, 1.25) {};
		\node [style=none] (8) at (0, -2.5) {};
		\node [style=none] (9) at (1, -1.25) {};
		\node [style=none] (10) at (0, 0) {};
		\node [style=none] (11) at (1, -0.25) {\ldots};
		\node [style=none] (12) at (1, -2.25) {\ldots};
		\node [style=white phase  dot] (13) at (1, -1.25) {$\beta$};
		\node [style=none] (14) at (2, -2.5) {};
		\node [style=none] (15) at (2, 0) {};
		\node [style=none] (16) at (-2, -2.5) {};
		\node [style=none] (17) at (2, 2.5) {};
	\end{pgfonlayer}
	\begin{pgfonlayer}{edgelayer}
		\draw [style=swap, in=135, out=-90, looseness=1.00] (2.center) to (6);
		\draw [style=swap, in=45, out=-90, looseness=1.00] (3.center) to (6);
		\draw [style=swap, in=90, out=-45, looseness=1.00] (6) to (0.center);
		\draw [style=swap, in=90, out=-135, looseness=1.00] (6) to (1.center);
		\draw [style=swap, in=135, out=-90, looseness=1.00] (10.center) to (13);
		\draw [style=swap, in=45, out=-90, looseness=1.00] (15.center) to (13);
		\draw [style=swap, in=90, out=-45, looseness=1.00] (13) to (14.center);
		\draw [style=swap, in=90, out=-135, looseness=1.00] (13) to (8.center);
		\draw [style=swap] (17.center) to (15.center);
		\draw [style=swap] (1.center) to (16.center);
	\end{pgfonlayer}
\end{tikzpicture}\ \ =\ \ \begin{tikzpicture}[scale=0.80]
	\begin{pgfonlayer}{nodelayer}
		\node [style=none] (0) at (1.25, -1.5) {};
		\node [style=none] (1) at (-1.25, -1.5) {};
		\node [style=none] (2) at (-1.25, 1.5) {};
		\node [style=none] (3) at (1.25, 1.5) {};
		\node [style=none] (4) at (0, 1.25) {\ldots};
		\node [style=none] (5) at (0, -1) {\ldots};
		\node [style=white phase dot] (6) at (0, 0) {$\alpha + \beta$};
		\node [style=none] (7) at (0, 0) {};
	\end{pgfonlayer}
	\begin{pgfonlayer}{edgelayer}
		\draw [style=swap, in=135, out=-90, looseness=1.00] (2.center) to (6);
		\draw [style=swap, in=45, out=-90, looseness=1.00] (3.center) to (6);
		\draw [style=swap, in=90, out=-45, looseness=1.00] (6) to (0.center);
		\draw [style=swap, in=90, out=-135, looseness=1.00] (6) to (1.center);
	\end{pgfonlayer}
\end{tikzpicture}
\qquad\qquad\qquad
\begin{tikzpicture}[scale=0.80]
	\begin{pgfonlayer}{nodelayer}
		\node [style=white dot] (0) at (-6.25, -1) {};
		\node [style=none] (1) at (-6.25, -2) {};
		\node [style=gray dot] (2) at (-6.25, 1) {};
		\node [style=none] (3) at (-6.25, 2) {};
		\node [style=none] (4) at (-4.5, 0) {$=$};
		\node [style=gray dot] (5) at (-3, 1) {};
		\node [style=none] (6) at (-3, 2) {};
		\node [style=none] (7) at (-3, -2) {};
		\node [style=white dot] (8) at (-3, -1) {};
		\node [style=gray dot] (9) at (3.5, 1) {}; 
		\node [style=none] (10) at (3.5, 2) {};
		\node [style=none] (11) at (3.5, -2) {};
		\node [style=white dot] (12) at (3.5, -1) {};
		\node [style=white dot] (13) at (5, -1) {};
		\node [style=gray dot] (14) at (5, 1) {};
		\node [style=none] (15) at (5, -2) {};
		\node [style=none] (16) at (5, 2) {};
		\node [style=none] (17) at (6.75, 0) {$=$};
		\node [style=white dot] (18) at (8.75, 1) {};
		\node [style=none] (19) at (9.5, 2) {};
		\node [style=none] (20) at (9.5, -2) {};
		\node [style=gray dot] (21) at (8.75, -1) {};
		\node [style=none] (22) at (8, -2) {};
		\node [style=none] (23) at (8, 2) {};
	\end{pgfonlayer}
	\begin{pgfonlayer}{edgelayer}
		\draw [style=swap] (0) to (1.center);
		\draw [style=swap, bend right=45, looseness=1.00] (0) to (2);
		\draw [style=swap] (2) to (3.center);
		\draw [style=swap, bend right=45, looseness=1.00] (2) to (0);
		\draw [style=swap] (8) to (7.center);
		\draw [style=swap] (5) to (6.center);
		\draw [style=swap] (12) to (11.center);
		\draw [style=swap] (9) to (10.center);
		\draw [style=swap, bend right=45, looseness=1.00] (9) to (12);
		\draw [style=swap] (13) to (15.center);
		\draw [style=swap] (14) to (16.center);
		\draw [style=swap, bend left=45, looseness=1.00] (14) to (13);
		\draw [style=swap] (9) to (13);
		\draw [style=swap] (14) to (12);
		\draw [style=swap] (18) to (19.center);
		\draw [style=swap] (21) to (20.center);
		\draw [style=swap] (23.center) to (18);
		\draw [style=swap] (18) to (21);
		\draw [style=swap] (21) to (22.center);
	\end{pgfonlayer}
\end{tikzpicture}
\eeq
So we are genuinely dealing with an entirely pictorial language, and in fact, recently it was shown in \cite{hadzihasanovic2018two} that all the equations that can be derived using the usual quantum mechanical formalism, can also be derived using only pictures!

\bM The key part of the passage from a sentence diagram representing grammatical structure like (\ref{eq:tv2wg}) to a quantum circuit uses the ZX-calculus in a fundamental way. For example, the fact that the CNOT-gate arises by sticking together two spiders as follows:
\beq
\begin{tikzpicture}[dotpic]
	\begin{pgfonlayer}{nodelayer}
		\node [style=none] (0) at (-1.5, -0.25) {};
		\node [style=white dot] (1) at (-0.75, -1.25) {};
		\node [style=none] (2) at (-0.75, -2.5) {};
		\node [style=none] (3) at (1.5, 0.25) {};
		\node [style=none] (4) at (0.75, 2.5) {};
		\node [style=gray dot] (5) at (0.75, 1.25) {};
		\node [style=none] (6) at (0, -0.25) {};
		\node [style=none] (7) at (0, 0.25) {}; 
		\node [style=none] (8) at (1.5, -2.5) {};
		\node [style=none] (9) at (-1.5, 2.5) {};
		\node [style=none] (10) at (-1.75, -0.25) {};
		\node [style=none] (11) at (0.25, -0.25) {};
		\node [style=none] (12) at (0.25, -2.25) {};
		\node [style=none] (13) at (-1.75, -2.25) {};
		\node [style=none] (14) at (1.75, 0.25) {};
		\node [style=none] (15) at (-0.25, 0.25) {};
		\node [style=none] (16) at (1.75, 2.25) {};
		\node [style=none] (17) at (-0.25, 2.25) {}; 
	\end{pgfonlayer}
	\begin{pgfonlayer}{edgelayer}
		\draw [in=-90, out=135, looseness=1.00] (1) to (0.center);
		\draw (2.center) to (1);
		\draw (5) to (4.center);
		\draw [in=-30, out=90, looseness=1.00] (3.center) to (5);
		\draw [style=swap, in=-90, out=45, looseness=1.00] (1) to (6.center);
		\draw [style=swap, in=-150, out=90, looseness=1.00] (7.center) to (5);
		\draw [style=swap] (7.center) to (6.center);
		\draw [style=swap] (3.center) to (8.center);
		\draw [style=swap] (9.center) to (0.center);
		\draw [style=dashed] (17.center) to (16.center);
		\draw [style=dashed] (16.center) to (14.center);
		\draw [style=dashed] (14.center) to (15.center);
		\draw [style=dashed] (17.center) to (15.center);
		\draw [style=dashed] (10.center) to (13.center);
		\draw [style=dashed] (13.center) to (12.center);
		\draw [style=dashed] (12.center) to (11.center);
		\draw [style=dashed] (11.center) to (10.center);
	\end{pgfonlayer}
\end{tikzpicture}
\eeq
is essential.  Using this decomposition, the circuit arises as follows from (\ref{eq:tv2wg}):
\beq
\begin{tikzpicture}[scale=0.80]
	\begin{pgfonlayer}{nodelayer}
		\node [style=none] (0) at (-11.25, -0.5) {};
		\node [style=langstate] (1) at (-11.25, 0.5) {\!\!\!{\tt Alice}\!\!\!};
		\node [style=langstate] (2) at (-2.75, 0.5) {\!\!\!{\tt Bob}\!\!\!};
		\node [style=none] (3) at (-2.75, -0.5) {};
		\node [style=none] (4) at (-5.25, -0.5) {};
		\node [style=none] (5) at (-8.75, -0.5) {};
		\node [style=none] (6) at (-2.75, -0.5) {};
		\node [style=none] (7) at (-8.75, -0.5) {};
		\node [style=none] (8) at (-5.25, -0.5) {};
		\node [style=none] (9) at (-5.25, -0.5) {};
		\node [style=none] (10) at (-8.75, -0.5) {};
		\node [style=none] (11) at (-11.25, -0.5) {};
		\node [style=none] (12) at (-8.75, -0.5) {};
		\node [style=none] (13) at (-6.75, -0.5) {};
		\node [style=none] (14) at (-8, 0.25) {};
		\node [style=none] (15) at (-6, 1) {};
		\node [style=langstate] (16) at (-7, 1.25) {\tt\!*hates*\!};
		\node [style=none] (17) at (-6, 0.25) {};
		\node [style=none] (18) at (-7.25, -0.5) {};
		\node [style=none] (19) at (-5.25, -0.5) {};
		\node [style=white dot] (20) at (-6, 0.25) {};
		\node [style=white dot] (21) at (-8, 0.25) {};
		\node [style=none] (22) at (-8.75, -0.5) {};
		\node [style=none] (23) at (-8, 1) {};
		\node [style=none] (24) at (-7.25, -2) {};
		\node [style=none] (25) at (-6.75, -2) {};
		\node [style=none] (26) at (-7, 2) {};
		\node [style=none] (27) at (-9.25, -0.5) {};
		\node [style=none] (28) at (-9.25, 1.75) {};
		\node [style=none] (29) at (-4.75, -0.5) {};
		\node [style=none] (30) at (-4.75, 1.75) {};
		\node [style=none] (31) at (0, 0) {$=$};
		\node [style=none] (32) at (2.75, -0.25) {};
		\node [style=none] (33) at (8, -1.25) {};
		\node [style=gray dot] (34) at (8, -0.25) {};
		\node [style=none] (35) at (6, -0.25) {};
		\node [style=none] (36) at (11.25, -0.25) {};
		\node [style=none] (37) at (6, 0.75) {};
		\node [style=gray dot] (38) at (6, -1.25) {};
		\node [style=none] (39) at (2.75, 0.75) {};
		\node [style=white dot] (40) at (2.75, -0.25) {};
		\node [style=gray dot] (41) at (6, -0.25) {};
		\node [style=none] (42) at (2.75, -1.75) {};
		\node [style=none] (43) at (11.25, -1.75) {};
		\node [style=langstate] (44) at (7, 1) {\tt\!*hates*\!};
		\node [style=none] (45) at (11.25, 0.75) {};
		\node [style=none] (46) at (2.75, -0.25) {};
		\node [style=none] (47) at (2.75, 0.75) {};
		\node [style=none] (48) at (6, -1.25) {};
		\node [style=white dot] (49) at (11.25, -0.25) {}; 
		\node [style=none] (50) at (8, 0.75) {};
		\node [style=none] (51) at (11.25, 0.75) {};
		\node [style=langstate] (52) at (2.75, 1) {\!\!\!{\tt Alice}\!\!\!};
		\node [style=none] (53) at (11.25, 0.75) {};
		\node [style=langstate] (54) at (11.25, 1) {\!\!\!{\tt Bob}\!\!\!};
		\node [style=none] (55) at (11.25, -0.25) {};
		\node [style=gray dot] (56) at (8, -1.25) {};
		\node [style=none] (57) at (2.75, 0.75) {};
		\node [style=none] (58) at (8, -0.25) {};
	\end{pgfonlayer}
	\begin{pgfonlayer}{edgelayer}
		\draw [style=swap, bend right=90, looseness=1.50] (8.center) to (6.center);
		\draw [style=swap, bend left=90, looseness=1.50] (7.center) to (11.center);
		\draw [style=swap] (14.center) to (23.center);
		\draw [style=swap, in=90, out=-15, looseness=1.00] (21) to (18.center);
		\draw [style=swap, in=90, out=-165, looseness=1.00] (21) to (22.center);
		\draw [style=swap] (17.center) to (15.center);
		\draw [style=swap, in=90, out=-15, looseness=1.00] (20) to (19.center);
		\draw [style=swap, in=90, out=-165, looseness=1.00] (20) to (13.center);
		\draw [style=swap] (18.center) to (24.center);
		\draw [style=swap] (13.center) to (25.center);
		\draw [style=swap, dashed] (28.center) to (27.center);
		\draw [style=swap, dashed] (27.center) to (29.center);
		\draw [style=swap, dashed] (29.center) to (30.center);
		\draw [style=swap, dashed] (26.center) to (28.center);
		\draw [style=swap, dashed] (26.center) to (30.center);
		\draw [style=swap] (1) to (0.center);
		\draw [style=swap] (2) to (3.center);
		\draw [style=swap] (32.center) to (39.center);
		\draw [style=swap] (46.center) to (42.center);
		\draw [style=swap, in=0, out=180, looseness=1.25] (35.center) to (32.center);
		\draw [style=swap] (37.center) to (41);
		\draw [style=swap] (41) to (38);
		\draw [style=swap] (55.center) to (51.center);
		\draw [style=swap] (36.center) to (43.center);
		\draw [style=swap, in=180, out=0, looseness=1.25] (58.center) to (55.center);
		\draw [style=swap] (50.center) to (34);
		\draw [style=swap] (34) to (56);
	\end{pgfonlayer}
\end{tikzpicture}
\eeq
We could then also decide to reduce the number of qubits as follows:
\beq
\begin{tikzpicture}[scale=0.80]
	\begin{pgfonlayer}{nodelayer}
		\node [style=none] (0) at (0, 0) {$=$};
		\node [style=langstate] (1) at (-12.25, 1) {\!\!\!{\tt Alice}\!\!\!};
		\node [style=white dot] (2) at (-12.25, -0.25) {};
		\node [style=none] (3) at (-12.25, -1.75) {};
		\node [style=none] (4) at (-8.5, -0.25) {};
		\node [style=none] (5) at (-12.25, -0.25) {};
		\node [style=none] (6) at (-12.25, 0.75) {};
		\node [style=none] (7) at (-12.25, -0.25) {};
		\node [style=none] (8) at (-12.25, 0.75) {};
		\node [style=none] (9) at (-12.25, 0.75) {};
		\node [style=langstate] (10) at (-2.75, 1) {\!\!\!{\tt Bob}\!\!\!};
		\node [style=gray dot] (11) at (-8.5, -0.25) {};
		\node [style=gray dot] (12) at (-8.5, -1.25) {};
		\node [style=none] (13) at (-8.5, -1.25) {};
		\node [style=none] (14) at (-8.5, 0.75) {};
		\node [style=gray dot] (15) at (-6, -1.25) {};
		\node [style=none] (16) at (-6, 0.75) {};
		\node [style=white dot] (17) at (-2.75, -0.25) {};
		\node [style=none] (18) at (-2.75, 0.75) {};
		\node [style=none] (19) at (-2.75, 0.75) {};
		\node [style=none] (20) at (-2.75, -0.25) {};
		\node [style=none] (21) at (-6, -1.25) {};
		\node [style=gray dot] (22) at (-6, -0.25) {};
		\node [style=none] (23) at (-2.75, 0.75) {};
		\node [style=none] (24) at (-2.75, -1.75) {};
		\node [style=none] (25) at (-6, -0.25) {};
		\node [style=none] (26) at (-2.75, -0.25) {};
		\node [style=langbox] (27) at (-8.5, 1) {\!\!\!{\tt *hates*}\!\!\!};
		\node [style=none] (28) at (-6, 0.75) {};
		\node [style=none] (29) at (-8.5, 1.25) {};
		\node [style=none] (30) at (-6, 1.25) {};
		\node [style=none] (31) at (-8.5, 0.75) {};
		\node [style=none] (32) at (9.25, 2.25) {};
		\node [style=gray dot] (33) at (6, 1.25) {};
		\node [style=gray dot] (34) at (6, -2.25) {}; 
		\node [style=gray dot] (35) at (6, 2.5) {};
		\node [style=none] (36) at (6, -2.25) {};
		\node [style=langstate] (37) at (2.75, 2.5) {\!\!\!{\tt Alice}\!\!\!};
		\node [style=none] (38) at (9.25, 2.25) {};
		\node [style=none] (39) at (9.25, 1.25) {};
		\node [style=none] (40) at (9.25, 2.25) {};
		\node [style=none] (41) at (2.75, -1.25) {};
		\node [style=gray dot] (42) at (6, -1.25) {};
		\node [style=none] (43) at (2.75, 2.25) {};
		\node [style=none] (44) at (6, 0.25) {};
		\node [style=none] (45) at (6, -0.25) {};
		\node [style=none] (46) at (6, 1.25) {}; 
		\node [style=none] (47) at (9.25, 1.25) {};
		\node [style=langbox] (48) at (6, 0) {\!\!\!{\tt *hates*}\!\!\!};
		\node [style=langstate] (49) at (9.25, 2.5) {\!\!\!{\tt Bob}\!\!\!};
		\node [style=white dot] (50) at (2.75, -1.25) {};
		\node [style=none] (51) at (2.75, 2.25) {}; 
		\node [style=none] (52) at (2.75, 2.25) {};
		\node [style=none] (53) at (6, -1.25) {};
		\node [style=none] (54) at (2.75, -2.75) {};
		\node [style=none] (55) at (2.75, -1.25) {};
		\node [style=white dot] (56) at (9.25, 1.25) {}; 
		\node [style=none] (57) at (6, -0.25) {};
		\node [style=none] (58) at (9.25, -2.75) {};
	\end{pgfonlayer}
	\begin{pgfonlayer}{edgelayer}
		\draw [style=swap] (7.center) to (6.center);
		\draw [style=swap] (5.center) to (3.center);
		\draw [style=swap, in=0, out=180, looseness=1.25] (4.center) to (7.center);
		\draw [style=swap] (14.center) to (11);
		\draw [style=swap] (11) to (12);
		\draw [style=swap] (26.center) to (18.center);
		\draw [style=swap] (20.center) to (24.center);
		\draw [style=swap, in=180, out=0, looseness=1.25] (25.center) to (26.center);
		\draw [style=swap] (16.center) to (22);
		\draw [style=swap] (22) to (15);
		\draw [style=swap] (30.center) to (28.center);
		\draw [style=swap, bend left=90, looseness=1.50] (29.center) to (30.center);
		\draw [style=swap] (41.center) to (52.center);
		\draw [style=swap] (55.center) to (54.center);
		\draw [style=swap, in=0, out=180, looseness=1.25] (53.center) to (41.center);
		\draw [style=swap] (45.center) to (42);
		\draw [style=swap] (42) to (34);
		\draw [style=swap] (39.center) to (38.center);
		\draw [style=swap] (47.center) to (58.center);
		\draw [style=swap, in=180, out=0, looseness=1.25] (46.center) to (39.center);
		\draw [style=swap] (33) to (35);
		\draw [style=swap] (33) to (44.center);
	\end{pgfonlayer}
\end{tikzpicture}
\eeq
From this (\ref{tv17more}) arises when parametrising the word meanings by gates.  A detailed discussion of the entire passage is in \cite{QNLP-foundations}.\e

In the form (\ref{tv17more}), natural language can be implemented on NISQ devices, and of course, will still work well as these devices scale in terms of size and performance.  Crucially, our solution provides a way forward in the absence of QRAM. By employing quantum machine learning we do not directly encode the meanings of words, but instead quantum gates---those in the  circuit (\ref{tv17more}) carrying Greek letters---learn their meanings directly from text \cite{QPL-QNLP}. By way of analogy with classical machine learning, in quantum machine learning  we can indeed use quantum circuits instead of classical neural networks in order to learn patterns from data \cite{ma2019variational, benedetti2019parameterized}. Interestingly, neural network architectures are the state-of-the-art in classical NLP, but the majority of methods do not take advantage of grammatical structures. In contrast, we saw that our approach to natural language naturally accommodates both grammar and meaning.

\begin{figure}
\centering
\epsfig{figure=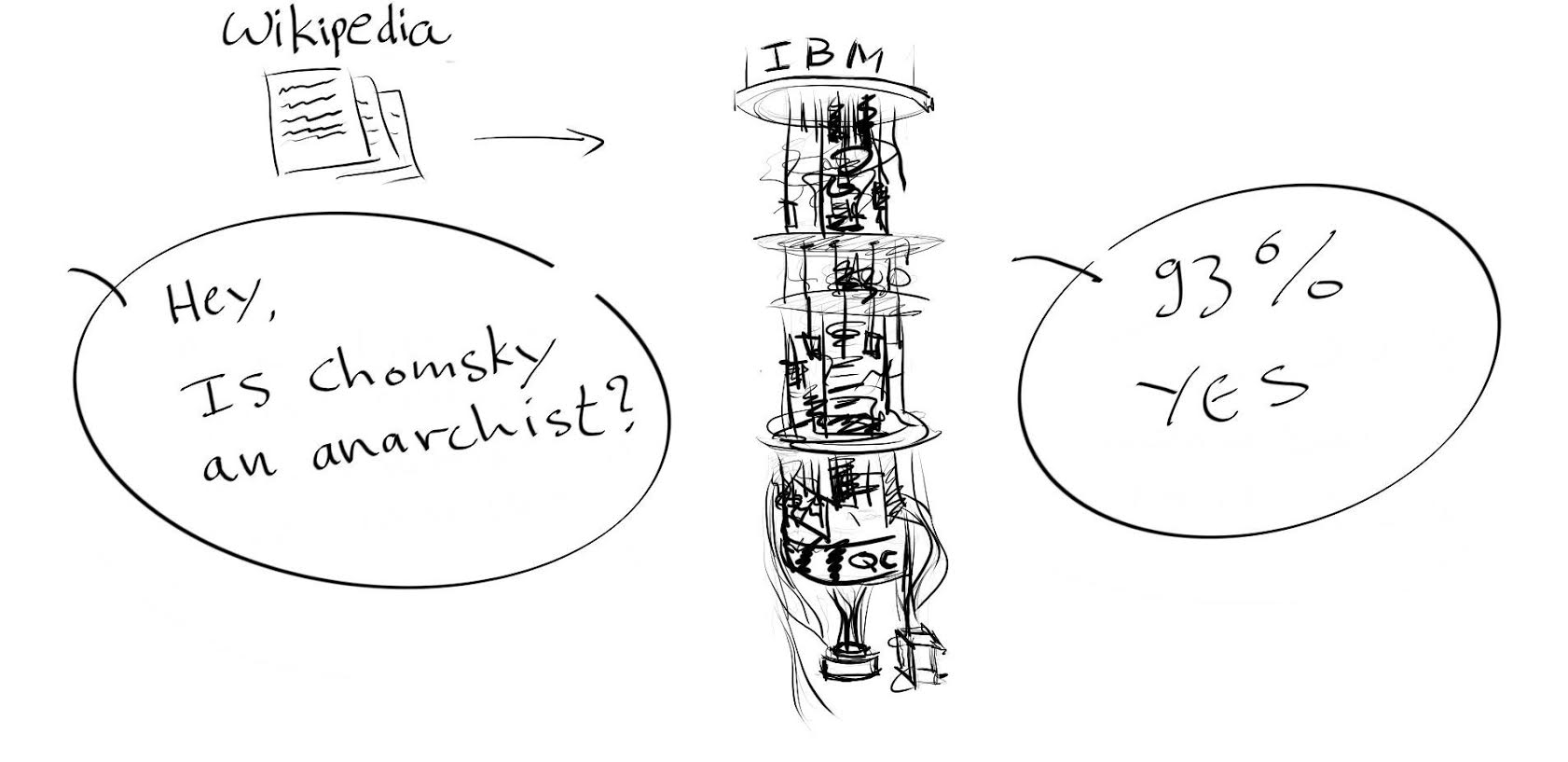,width=454pt}
\caption{A quantum computer putting an end to social media disputes.}\label{Kcart}
\end{figure}

Using our framework, once the meanings of words and phrases are encoded as quantum gates, we are able to encode the meaning of grammatical sentences on quantum hardware. Posing a question to the quantum computer, constructed by the vocabulary and grammar the quantum computer has learned, it returns the answer\bM---as illustrated in Figure \ref{Kcart}\e.\footnote{Earlier theoretical work by our team towards question-answering includes \cite{CDMT, de2019functorial}.}

Naturally, we next turned our attention toward the design and execution of an experiment that is non-trivial, not least of all since our design is predicated on the program being scalable. This means that the dimension of the meaning space grows significantly with the number of qubits available whilst the size of the circuits dictated by the grammar does not grow too large with the size of the sentence.
%Interestingly, we note further that QNLP on NISQ devices is a novel playground for quantifying the scalability of quantum machine learning algorithms and experimenting with various quantum meaning spaces. Being able to accommodate large chunks of text is crucial if we aspire to graduate from sentence to text, and we know that we have made the first step towards this vision.

\section{The actual experiment}

%Our experimental workflow is as follows. Let G be the grammar category, which is the mathematical model where grammar diagrams are generated. As stated above, a grammar diagram (or network) encodes the information flow of word-meanings in a grammatical sentence.
%In more detail, a diagram is none other than a grammatical and syntactic parsing of a sentence according to a specified grammar model. This diagram is then instantiated as a quantum circuit that lives in the category QCirc(?). Next, the meaning of the words is encoded in quantum states in such quantum circuits. Specifically, any state can be prepared from a classical reference state, and so by ?state? we refer to the circuit (or process) that prepares it. Then the composition of words in a sentence corresponds to composition of circuits representing words. This results in a circuit that prepares a state encoding the meaning of a sentence.
%Importantly, the circuits are parameterised by a set ?. In other words, these ?grammatical quantum circuits? are a family spanned by ?. By allowing circuits to depend on parameters we create the semantic space in which the meanings of the words, and consequently whole sentences, are encoded.
%Image for post

\begin{figure}
\centering
\epsfig{figure=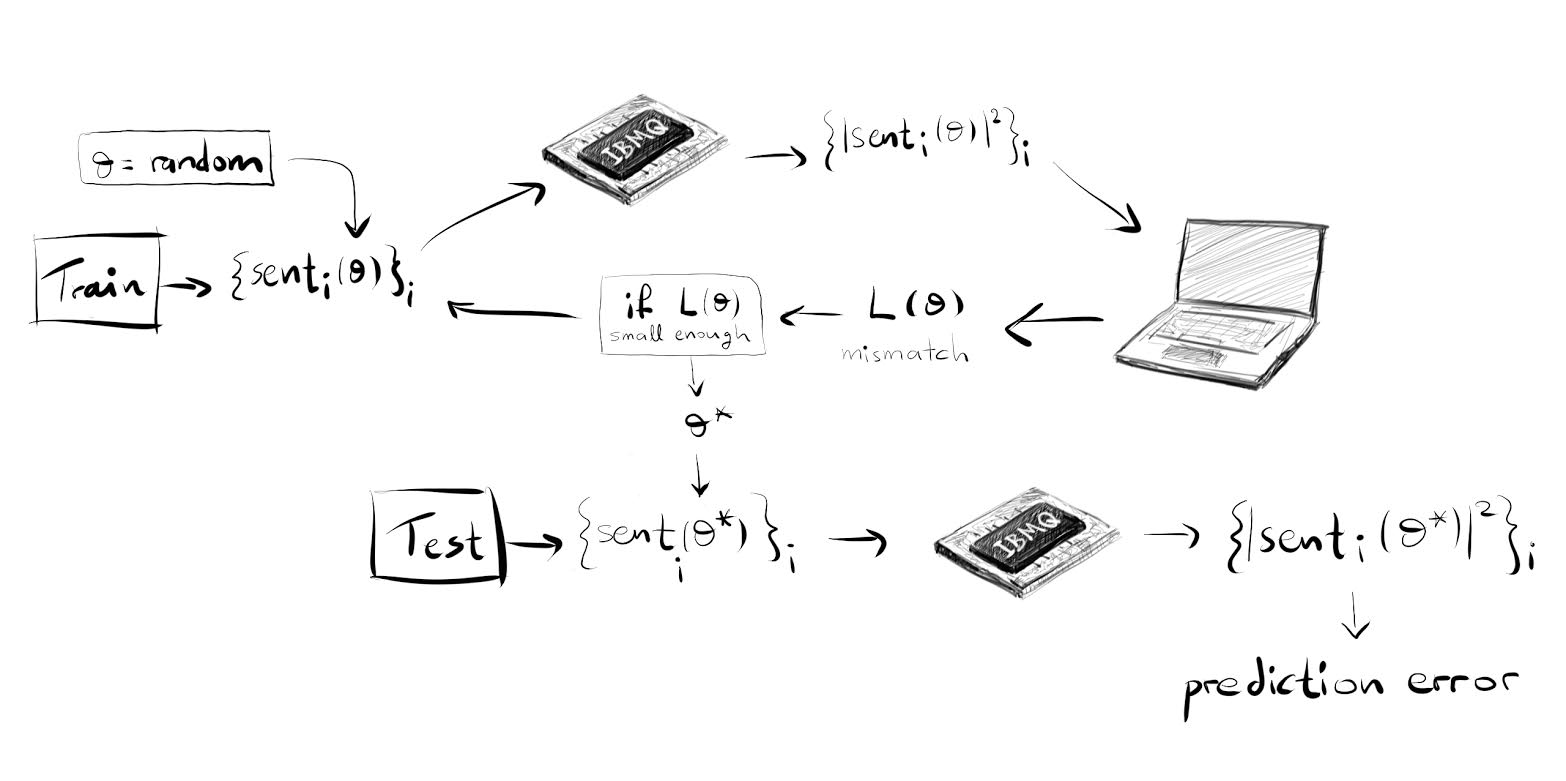,width=440pt}
\bM\caption{A schematic presentation of the actual experiment on IBM quantum hardware, depicting both the training and testing parts, with $\theta= \{\alpha_A, \beta_A,\alpha_B, \beta_B, \alpha, \beta, \ldots\}$.}\label{Pipe}\e
\end{figure}

Once a quantum circuit is created from a sentence, it needs to be evaluated in order to compute the meaning of that sentence. We may choose to perform this evaluation on a classical computer, where we employ state-of-the-art methods for performing the costly task of multiplying exponentially big matrices, or, we may choose to implement the circuit on a quantum computer.  This is of course what we decided to do.  \bM A schematic presentation of the experiment on IBM quantum hardware is in Figure \ref{Pipe}. \e

As we see in the quantum circuit (\ref{tv17more}),  each of the parts of speech (subject, object, verb) is in the quantum circuit a function of some parameters. For example, there are sets of parameter values $\alpha_A, \beta_A,\alpha_B, \beta_B, \alpha, \beta, \ldots$ such that:
\beqa
{\tt subject}(\alpha_A, \beta_A)&=&{\tt Alice}\\
{\tt object}(\alpha_B, \beta_B)&=&{\tt Bob}\\
{\tt verb}(\alpha, \beta, \ldots)&=&{\tt hates}
\eeqa
The values are determined empirically by a text corpus and are then used to answer questions about the corpus. In order to ensure that our experiment can be executed effectively on near-term NISQ devices, but at the same time be complex enough to be interesting, we chose a vocabulary of a few words, for example:
\[
\left\{{\tt Alice}, {\tt Bob}, {\tt loves}, {\tt hates}, {\tt rich}, {\tt silly}\right\}
\]
and generated not some, but all grammatical sentences from their combinations. From these sentences, we created their corresponding parameterised circuits. Moreover, we interpret the language diagrams such that the sentence space is one-dimensional, i.e.~just a number indicating the truth-value of the sentence:
\[
0\leq {\tt sentence}(\alpha_A, \beta_A,\alpha_B, \beta_B, \alpha, \beta, \ldots)\leq 1
\]
A value close to 1 represents ``$true$" and a value close to 0 indicates ``$false$".

The labeled toy corpus would look like:
\[
Corpus= \{({\tt Alice\ loves\ Bob}, false), ({\tt Bob\ is\ silly}, true), ({\tt Alice\ is\ rich}, true), ... \}
\]
Now that we have our corpus of sentences we split the corpus:
\[
Corpus=Train\cup Test
\]
in a training set $Train$ and a test set $Test$. Sentences in the training set $Train$ are used to do supervised quantum machine learning in order to learn the parameters that result in the correct measurement of the truth-labels. In this way, the parameters for the circuits that prepare the meaning states for nouns $\{{\tt Alice}, {\tt Bob}\}$, verbs $\{{\tt is}, {\tt loves}, {\tt hates}\}$, and adjectives $\{{\tt rich}, {\tt silly}\}$, are learned.

The scheme for learning the parameters for words in sentences in the training set is as follows. The circuit of a sentence in the training set is evaluated for the current set of parameters on the quantum computer. By sampling measurement outcomes we estimate:
\[
\left|{\tt sentence}(\alpha_A, \beta_A,\alpha_B, \beta_B, \alpha, \beta, \ldots)\right|^2
\]
This number is read by a classical computer that checks how well this matches the desired truth label of the sentence. If there is a mismatch, our system updates the parameters so that the quantum computer may evaluate an updated circuit. Iterating this procedure until the parameters converge and all truth labels are reproduced for the sentences in the training set.

After training, sentences in $Test$ are used to estimate how well the truth labels of new sentences, i.e.~not in $Train$, are inferred. These new sentences share the same vocabulary as the sentences used for training, but they are grammatically and semantically different.

Note that finding the optimal sequence of updates is, in general, a so-called `hard optimization problem', so it is important that our quantum meaning space is well designed and allows the learning to be as tractable as possible. This design feature is critical.

With this learning framework, we can now ask questions of the quantum computer, as long as the questions are grammatical sentences expressed in terms of the vocabulary and grammar used during training. We are pleased to add that in our experiment, questions can, in fact, be posed as compositions of already learned sentences. For example, we can use the relative pronoun {\tt who} \cite{FrobMeanI} (which we model in terms of CNOT gates within a quantum circuit) and ask:
\[
{\tt Does\ Bob\ who\ is\ silly\ love\ Alice\ who\ is\ rich?}
\]
This is the same as asking whether:
\beq
\begin{tikzpicture}[scale=0.80]
	\begin{pgfonlayer}{nodelayer}
		\node [style=none] (0) at (-5.75, 0) {};
		\node [style=langstate] (1) at (-12.5, 0.5) {\!\!\!{\tt Bob}\!\!\!};
		\node [style=none] (2) at (-2.5, 0) {};
		\node [style=none] (3) at (-9.25, -0.25) {};
		\node [style=none] (4) at (-7.25, 0) {};
		\node [style=none] (5) at (-5.75, 0.25) {};
		\node [style=langstate] (6) at (-5.25, 0.5) {\!\!\!{\tt is}\!\!\!};
		\node [style=none] (7) at (-9.25, 0) {};
		\node [style=none] (8) at (-8.25, 0) {};
		\node [style=none] (9) at (-4.75, 0) {};
		\node [style=none] (10) at (-5.25, 0) {};
		\node [style=none] (11) at (-7.25, 0) {};
		\node [style=none] (12) at (-5.75, 0) {};
		\node [style=none] (13) at (-5.25, 0) {};
		\node [style=none] (14) at (-10.25, 0) {};
		\node [style=none] (15) at (-4.75, 0.25) {};
		\node [style=none] (16) at (-10.25, 0) {};
		\node [style=none] (17) at (-12.5, 0) {};
		\node [style=langstate] (18) at (-2.5, 0.5) {\!\!\!{\tt silly}\!\!\!};
		\node [style=white ddot] (19) at (-8.25, 0.25) {};
		\node [style=white dot] (20) at (-8.75, 1) {};
		\node [style=none] (21) at (-5.25, 0.25) {};
		\node [style=none] (22) at (-2.5, 0.25) {};
		\node [style=none] (23) at (-12.5, 0.25) {};
		\node [style=none] (24) at (-8.25, 0) {};
		\node [style=none] (25) at (-8.25, 0.25) {};
		\node [style=none] (26) at (-6.75, 1.25) {};
		\node [style=none] (27) at (-10.75, 1.25) {};
		\node [style=none] (28) at (-8.75, 1.5) {};
		\node [style=none] (29) at (-10.75, -0.25) {};
		\node [style=none] (30) at (-6.75, -0.25) {};
		\node [style=none] (31) at (-8.75, 2) {\footnotesize\tt who};
		\node [style=none] (32) at (11.75, 0.25) {};
		\node [style=none] (33) at (11.25, 0) {};
		\node [style=none] (34) at (8.75, 0.25) {};
		\node [style=white dot] (35) at (8.25, 1) {};
		\node [style=none] (36) at (11.25, 0) {};
		\node [style=none] (37) at (9.75, 0) {};
		\node [style=none] (38) at (9.75, 0) {};
		\node [style=none] (39) at (10.25, -0.25) {};
		\node [style=none] (40) at (8.25, 1.5) {};
		\node [style=none] (41) at (8.75, 0) {};
		\node [style=none] (42) at (14.5, 0) {};
		\node [style=langstate] (43) at (11.75, 0.5) {\!\!\!{\tt is}\!\!\!};
		\node [style=none] (44) at (6.25, -0.25) {};
		\node [style=none] (45) at (11.25, 0.25) {};
		\node [style=none] (46) at (10.25, 1.25) {};
		\node [style=none] (47) at (6.75, 0) {};
		\node [style=white ddot] (48) at (8.75, 0.25) {};
		\node [style=none] (49) at (12.25, 0) {};
		\node [style=langstate] (50) at (14.5, 0.5) {\!\!\!{\tt rich}\!\!\!};
		\node [style=none] (51) at (14.5, 0.25) {};
		\node [style=none] (52) at (4.5, 0.25) {};
		\node [style=none] (53) at (6.75, 0) {};
		\node [style=none] (54) at (7.75, -0.25) {};
		\node [style=none] (55) at (12.25, 0.25) {};
		\node [style=none] (56) at (7.75, 0) {};
		\node [style=none] (57) at (8.75, 0) {};
		\node [style=langstate] (58) at (4.5, 0.5) {\!\!\!{\tt Alice}\!\!\!};
		\node [style=none] (59) at (8.25, 2) {\footnotesize\tt who};
		\node [style=none] (60) at (4.5, 0) {};
		\node [style=none] (61) at (11.75, 0) {};
		\node [style=none] (62) at (11.75, 0) {};
		\node [style=none] (63) at (6.25, 1.25) {};
		\node [style=none] (64) at (2, -0.25) {};
		\node [style=none] (65) at (0, -0.25) {};
		\node [style=none] (66) at (0, -0.25) {};
		\node [style=langstate] (67) at (1, 0.5) {\!\!\!{\tt loves}\!\!\!};
		\node [style=none] (68) at (1, 0) {};
		\node [style=none] (69) at (1, 0.25) {};
		\node [style=none] (70) at (0, 0.25) {};
		\node [style=none] (71) at (2, 0.25) {};
		\node [style=none] (72) at (1, -3) {};
	\end{pgfonlayer}
	\begin{pgfonlayer}{edgelayer} 
		\draw [style=swap] (9.center) to (15.center);
		\draw [style=swap, bend right=90, looseness=1.50] (9.center) to (2.center);
		\draw [style=swap] (12.center) to (5.center);
		\draw [style=swap] (3.center) to (7.center);
		\draw [style=swap, bend right=90, looseness=1.50] (11.center) to (0.center);
		\draw [style=boldedge, bend right=90, looseness=1.25] (8.center) to (10.center);
		\draw [style=swap, bend right=90, looseness=1.75] (17.center) to (16.center);
		\draw [style=swap, in=0, out=90, looseness=1.00] (4.center) to (20);
		\draw [style=swap, in=-135, out=90, looseness=1.00] (7.center) to (20);
		\draw [style=swap, in=180, out=90, looseness=1.00] (14.center) to (20);
		\draw [style=swap] (21.center) to (13.center);
		\draw [style=swap] (23.center) to (17.center);
		\draw [style=swap] (22.center) to (2.center);
		\draw [style=boldedge] (24.center) to (25.center);
		\draw [style=swap, dashed] (27.center) to (29.center);
		\draw [style=swap, dashed] (29.center) to (30.center);
		\draw [style=swap, dashed] (30.center) to (26.center);
		\draw [style=swap, dashed] (28.center) to (27.center);
		\draw [style=swap, dashed] (28.center) to (26.center);
		\draw [style=swap] (49.center) to (55.center);
		\draw [style=swap, bend right=90, looseness=1.50] (49.center) to (42.center);
		\draw [style=swap] (33.center) to (45.center);
		\draw [style=swap] (54.center) to (56.center);
		\draw [style=swap, bend right=90, looseness=1.50] (37.center) to (36.center);
		\draw [style=boldedge, bend right=90, looseness=1.25] (57.center) to (61.center);
		\draw [style=swap, bend right=90, looseness=1.75] (60.center) to (53.center);
		\draw [style=swap, in=0, out=90, looseness=1.00] (38.center) to (35);
		\draw [style=swap, in=-135, out=90, looseness=1.00] (56.center) to (35);
		\draw [style=swap, in=180, out=90, looseness=1.00] (47.center) to (35);
		\draw [style=swap] (32.center) to (62.center);
		\draw [style=swap] (52.center) to (60.center);
		\draw [style=swap] (51.center) to (42.center);
		\draw [style=boldedge] (41.center) to (34.center);
		\draw [style=swap, dashed] (63.center) to (44.center);
		\draw [style=swap, dashed] (44.center) to (39.center);
		\draw [style=swap, dashed] (39.center) to (46.center);
		\draw [style=swap, dashed] (40.center) to (63.center);
		\draw [style=swap, dashed] (40.center) to (46.center);
		\draw [style=swap] (64.center) to (71.center);
		\draw [style=swap] (66.center) to (70.center);
		\draw [style=swap] (69.center) to (68.center);
		\draw [style=boldedge] (69.center) to (72.center);
		\draw [style=swap, bend right=90, looseness=0.75] (3.center) to (65.center);
		\draw [style=swap, bend right=90, looseness=1.00] (64.center) to (54.center);
	\end{pgfonlayer}
\end{tikzpicture}
\eeq
is $true$. This amounts to evaluating a bigger quantum circuit than the one that the model has been trained on. However, because the model was trained on the same grammar and vocabulary as used to express the question, we get the expected truth label, false in this case.
%In particular, the truth label is obtained by estimating the value of the circuit, depicted below, by sampling measurement outcomes for a large number of runs of the quantum circuit corresponding to the above sentence:
%Image for post
%The transformation of the question-sentence to a statement-sentence that can be mapped to a quantum circuit and evaluated on a quantum machine is done at the level of grammar diagrams by suitable transformations.?

One critical  enabling feature of our experiment is making computers understand the pictures that we have been drawing.  For this we used GdF and AT's DisCoPy Python library \cite{DisCoPy}.  Another critical feature is effective and efficient compiling and optimization. In order to run a circuit on a quantum device, it needed to be compiled. Compiling entails morphing the circuit such that quantum operations are expressed in terms of device-native operations, as well as accommodating for the quantum processor's restricted connectivity. For this task, we used Cambridge Quantum Computing's quantum software development platform, t$|$ket$\rangle$ \cite{sivarajah2020t} which again crucially makes use of ZX-calculus.

\section{QNLP}

We've done it, it exists, so let's give the kid a name: `quantum natural language processing', or QNLP in short.  A much more detailed account on QNLP as well as the data for a larger-scale experiment can be found in \cite{QNLP-foundations, Nature, QNLPPlus100}.  With the successful execution of QNLP on quantum hardware, and the great promise for the future,  Cambridge Quantum Computing has now resourced an Oxford based team dedicated to QNLP, and structure-driven quantum AI more generally, evidently including each of us.   Let's single out a few special features of QNLP.

Firstly, as already indicated above, QNLP is quantum-native.  What we mean by that is that the model of natural language that we employ is a quantum model \cite{teleling}.  Not just because we think that having a quantum model is particularly cool, but because it was effectively the case that the most economic manner for bringing language meaning and grammatical structure together was provided by  the `categorical quantum mechanics' framework \cite{AC1, Kindergarten, CKbook}.\footnote{Categorical quantum mechanics is really just a fancy name for doing quantum mechanics in terms of pictures.}  The immediate consequence of this is that when we try to `simulate' QNLP on a classical computer, it would be exponentially expensive, just as it is the case for simulation quantum systems on a classical computer.  Hence, QNLP truly loves being on quantum hardware.

Secondly, it is fair to say that it is meaning-aware, at the very least, it is definitely more meaning-aware than the current deep learning based NLP implementations.  Indeed, the language pictures clearly indicate how the meanings flow in a meaningful way.  We say a bit more about this below in Section \ref{sec:wholism}.

Thirdly, as already indicated in \cite{WillC}, there are many more advantages to using quantum computers, most notably, we get quantum speed-up for a large variety of NLP tasks, like question-answering.  This space of opportunities is in fact still largely unexploited, and the mission of our QNLP team is to do just that, in as many possible manners as we can.

\section{Beyond language}

We are now well on the way to make qubits speak.  But can we make them do other things as well?  We are pretty sure we can, a first thing could be to make qubits represent all of the inputs that we get through our fleshy physical embodiment, like taste, smell, vision, hearing.  The technical term for the kinds of spaces that represent these sensory modes is `conceptual spaces'.  They were introduced by Peter G{\"a}rdenfors \cite{gardenfors2, gardenfors}. There already are existing frameworks that represent the input of these senses in a manner that exactly matches our  language diagrams, and hence allow for these sensory inputs to interact with each other \cite{ConcSpacI}.

%But all of this is really only a first step towards to a broader conception of quantum AI, in line with our particular approach to natural language within the quantum realm.  That's really all we can say at this point, but stay tuned on this channel for what is coming next!

These conceptual spaces are convex, and fit well with some other work within the framework of our natural language diagrams, namely on the use of the quantum mechanical concept of density matrices, which also form convex spaces \cite{Ludwig, Barrett}. This additional convex structure allows one to bring many more linguistic features into play, like word ambiguity. For example, the word {\tt queen} is clearly ambiguous, as it can be a royal, a rock band, a bee, a chess piece, a playing card, etc.  In order to represent that ambiguity, we simply add all the different meanings in order to form a density matrix \cite{calco2015}:
\beqa
\rho_{\mbox{\scriptsize\tt queen}} &=& |\mbox{\tt queen-royal}\rangle\langle\mbox{\tt queen-royal}|   \\
&+&|\mbox{\tt queen-band}\rangle\langle\mbox{\tt queen-band}|   \\
&+&|\mbox{\tt queen-bee}\rangle\langle\mbox{\tt queen-bee}|   \\
&+&|\mbox{\tt queen-chess}\rangle\langle\mbox{\tt queen-chess}|   \\
&+&|\mbox{\tt queen-card}\rangle\langle\mbox{\tt queen-card}| \\
&+& \ldots
\eeqa

Alternatively, these density matrices can also be used to express hierarchical relationships between words \cite{bankova2019graded}, for example, {\tt lion}, {\tt tiger} and {\tt cheeta}, are examples of {\tt big cat}, which further generalises to {\tt mammal}, which generalises to {\tt vertebrate}, etc.  Using density matrices we encode this hierarchy as follows:
\beqa
\rho_{\mbox{\scriptsize\tt lion}} &=& |0\rangle\langle0|\\
\rho_{\mbox{\scriptsize\tt tiger}} &=& |+\rangle\langle+|\\
\rho_{\mbox{\scriptsize\tt cheeta}} &=& |-\rangle\langle-|\\
\rho_{\mbox{\scriptsize\tt big cat}} &=& |0\rangle\langle0| + |1\rangle\langle1|\\
\rho_{\mbox{\scriptsize\tt mammal}} &=& |0\rangle\langle0| + |1\rangle\langle1| + |2\rangle\langle2|\\
\rho_{\mbox{\scriptsize\tt vertebrate}} &=& |0\rangle\langle0| + |1\rangle\langle1| + |2\rangle\langle2| + |3\rangle\langle3|
\eeqa
In \cite{CoeckeMeich, MarthaDot, MarthaNeg, GemmaMartha, MarthaNeural} this use of density matrices in language is further elaborated upon.

\section{String diagrams are everywhere}

Sticking with the theme of big cats, what's the difference between a tiger and a lion? In particular, why does a tiger have stripes, while a lion doesn't.
%maybe more extreme difference, tiger vs polar bear.
\begin{figure}
\centering
\epsfig{figure=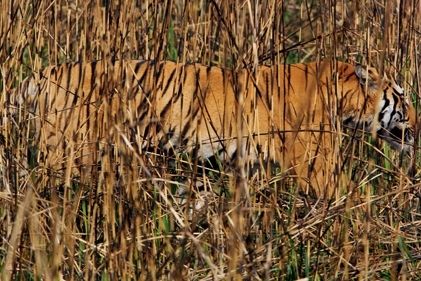,height=160pt}\epsfig{figure=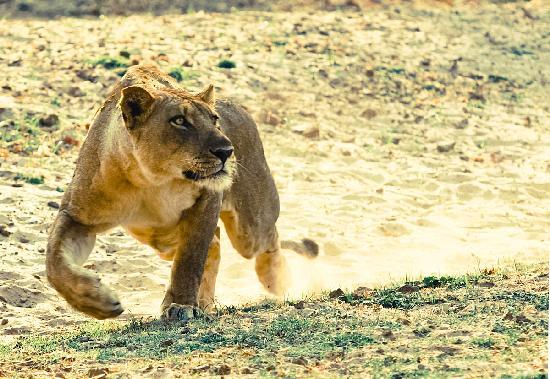,height=160pt}
\bM\caption{Two close relatives with very different coats.}\e
\end{figure}
A more traditional scientist will tell you that these days we  understand all of that perfectly well.  Before, if we dissected these animals, we found exactly the same organs.
%\begin{center}
%\epsfig{figure=figures/dissect,height=160pt}
%\end{center}
If we do some  smaller scale animal chopping, we end up with cells that again have exactly the same structure.  However, digging even further, and hitting the molecular level, we encounter DNA, and then we truly understand the difference.

Really? Some obscure humongously large code that is impossible to grasp by any human `explains' the difference?  It explains as much the difference between these two animals as the codes of two computer programs written in terms of 0's and 1's explain the difference between accounting software and the operating system of your smart phone.

That said, most of the exact sciences  have adopted this perspective of understanding things by breaking them down to smaller things, like elementary particles in particle physics, and the elements of a set in set-theory.  In contrast, in the arts and humanities  things really get their meanings by putting them in a context. For example, in the social sciences properties are mostly thought of in terms of behaviours, rather than by trying to build a map of the inside of someone's brain, and most modern art has very little meaning without a context.

We think that we can all agree that for our tiger and our lion, what explains the difference is really the hunting process which each of them needs to do in order to survive, and their respective coats provide the best camouflage for doing so, given they hunt in very different environments.  While of course these differences are encoded in the genes, it is the process of evolution that makes our tiger and our lion look as they do.
The role of DNA in explaining the stripes of tigers is similar to the role of our parameters $\alpha_A, \beta_A,\alpha_B, \dots$ in computing the meaning of sentences. Indeed, learning the optimal parameters is an evolution-like process that depends on the environment of the words, i.e. on the training data that we feed the machine.

So the two views explained above are really about either focussing on components vs.~context, and the latter is very much in line with Wittgenstein's ``meaning is use''.  We are saying all of this because the language diagrams that we saw above are the mathematical incarnation of this way of thinking, for example:
\beq
\begin{tikzpicture}[scale=0.80]
	\begin{pgfonlayer}{nodelayer}
		\node [style=none] (0) at (-9, 0.5) {};
		\node [style=langstate] (1) at (-9, 0.75) {\!\!\!{\tt tiger}\!\!\!};
		\node [style=langstate] (2) at (-1, 0.75) {\!\!\!{\tt deer}\!\!\!};
		\node [style=none] (3) at (-1, 0.5) {};
		\node [style=langstate] (4) at (-5, 0.75) {\ {\tt hunts}\ };
		\node [style=none] (5) at (1.5, 0.25) {};
		\node [style=none] (6) at (-3.5, 0.5) {};
		\node [style=none] (7) at (-5, 0.25) {};
		\node [style=none] (8) at (-6.5, 0.5) {};
		\node [style=none] (9) at (-1, 0.5) {};
		\node [style=none] (10) at (-6.5, 0.5) {};
		\node [style=none] (11) at (-3.5, 0.5) {};
		\node [style=none] (12) at (-3.5, 0.5) {};
		\node [style=none] (13) at (-6.5, 0.5) {};
		\node [style=none] (14) at (-9, 0.5) {};
		\node [style=none] (15) at (-6.5, 0.5) {};
		\node [style=none] (16) at (3.5, 0.5) {};
		\node [style=none] (17) at (3.5, 0.5) {};
		\node [style=none] (18) at (2.5, 0.5) {};
		\node [style=none] (19) at (2.5, -2) {};
		\node [style=langstate] (20) at (2.5, 0.75) {\ \ {\tt in}\ \ };
		\node [style=none] (21) at (1.5, 0.25) {};
		\node [style=none] (22) at (3.5, 0.5) {};
		\node [style=none] (23) at (1.5, 0.25) {};
		\node [style=none] (24) at (1.5, 0.25) {};
		\node [style=none] (25) at (1.5, 0.25) {};
		\node [style=langstate] (26) at (6.75, 0.75) {\!\!\!{\tt mangrove}\!\!\!};
		\node [style=none] (27) at (6, 0.5) {};
		\node [style=none] (28) at (7.5, 0.5) {};
		\node [style=none] (29) at (10.75, 0.5) {};
		\node [style=none] (30) at (10.75, 0.5) {};
		\node [style=langstate] (31) at (10.75, 0.75) {\!\!\!{\tt swamp}\!\!\!};
		\node [style=none] (32) at (7.5, 0.5) {};
		\node [style=none] (33) at (10.75, 0.5) {};
		\node [style=none] (34) at (10.75, 0.5) {};
		\node [style=none] (35) at (10.75, 0.5) {};
		\node [style=none] (36) at (10.75, 0.5) {};
		\node [style=none] (37) at (7.5, 0.5) {};
		\node [style=none] (38) at (-5, 0.5) {};
		\node [style=none] (39) at (1.5, 0.5) {};
	\end{pgfonlayer}
	\begin{pgfonlayer}{edgelayer}
		\draw [style=boldedge, bend right=90, looseness=1.00] (7.center) to (5.center);
		\draw [style=swap, bend right=90, looseness=1.50] (11.center) to (9.center);
		\draw [style=swap, bend left=90, looseness=1.50] (10.center) to (14.center);
		\draw [style=boldedge] (18.center) to (19.center);
		\draw [style=swap, bend left=90, looseness=1.50] (36.center) to (32.center);
		\draw [style=swap, bend right=90, looseness=1.75] (22.center) to (27.center);
		\draw [style=boldedge] (38.center) to (7.center);
		\draw [style=boldedge] (39.center) to (25.center);
	\end{pgfonlayer}
\end{tikzpicture}
\eeq
Indeed, this way of thinking requires a completely different kind of mathematics, be it in the form of our diagrams, or its symbolic counterpart, so-called monoidal categories.

This is not the place to go into this new kind of mathematics, but the interested reader may want to check out the book \cite{CKbook}, which is probably the most comprehensive account on these language diagrams, and their communality with quantum theory.
%\[
%\epsfig{figure=BookNew,width=260pt}
%\]
Shorter more informal accounts can be found in \cite{DBLP:books/daglib/p/Coecke17}.  For the braver reader, the symbolic counterpart of all of this exists within the realm of category theory, and tensor categories in particular \cite{CatsII, SelingerSurvey}.    Meanwhile, string diagrams have started to pop up in a wide variety of disciplines, for example, in computer science \cite{AbrRetracing, pavlovic2013monoidal}, where they have been for a while, in engineering \cite{SobocinskiSignal, BaezFongElec}, in economy and game theory \cite{ghani2018compositional, de2020functorial}, and even in cognition \cite{tull2020integrated, signorelli2020compositional}.

We believe that this is the kind of mathematics that would also serve the social sciences very well.  As it is quite new, most mathematics that one finds in the social sciences is statistics, which, in all fairness, is what one  uses in absence of anything better.

With  new mathematics also come new software tools.  For this we are using the new very flexible DisCoPy software \cite{DisCoPy} which we already mentioned above, and  continues  to be further developed by AT, GdF and collaborators. We're inviting everyone to play around with this toolbox, and even better, help us out further developing DisCoPy---we may even reward you with a pie. For ZX-calculus specifically there is now also the PyZX software \cite{kissinger2019pyzx}, which communicates directly with DisCoPy.  PyZX also helps us out with our QNLP experiments, by making the circuits that we feed into the quantum computer as simple as possible, a task for which ZX-calculus is now setting the state-of-the-art \cite{duncan2019graph, cowtan2019phase, de2020fast}.

\section{Compositionality}\label{sec:compose}

Above we saw how we can turn language diagrams into circuits.  A typical thing about circuits is, of course, that you can compose different circuits together for forming bigger circuits.  This is the principle of ``compositionality'', and in fact, string diagrams themselves are a result of this way of thinking. We can now, once we moved to circuits, compose several sentences together in order to produce larger text:
\beq
\begin{tikzpicture}[scale=0.80]
	\begin{pgfonlayer}{nodelayer}
		\node [style=langstate] (0) at (4.25, 0.75) {\tt\!likes\!};
		\node [style=white dot] (1) at (0, -2.5) {};
		\node [style=none] (2) at (0, -4) {};
		\node [style=none] (3) at (3.25, -2.5) {};
		\node [style=none] (4) at (0, -2.5) {};
		\node [style=none] (5) at (0, -1) {};
		\node [style=none] (6) at (0, -2.5) {};
		\node [style=langstate] (7) at (8.5, 0.75) {\!\!\!{\tt beer}\!\!\!};
		\node [style=gray dot] (8) at (3.25, -2.5) {};
		\node [style=gray dot] (9) at (3.25, -3.5) {};
		\node [style=none] (10) at (3.25, -3.5) {};
		\node [style=none] (11) at (3.25, 0.5) {};
		\node [style=gray dot] (12) at (5.25, -3.5) {};
		\node [style=none] (13) at (5.25, 0.5) {};
		\node [style=white dot] (14) at (8.5, -2.5) {};
		\node [style=none] (15) at (8.5, 0.5) {};
		\node [style=none] (16) at (8.5, 0.5) {};
		\node [style=none] (17) at (8.5, -2.5) {};
		\node [style=none] (18) at (5.25, -3.5) {};
		\node [style=gray dot] (19) at (5.25, -2.5) {};
		\node [style=none] (20) at (8.5, 0.5) {};
		\node [style=none] (21) at (8.5, -4) {};
		\node [style=none] (22) at (5.25, -2.5) {};
		\node [style=none] (23) at (8.5, -2.5) {};
		\node [style=none] (24) at (0, 0.5) {};
		\node [style=none] (25) at (-8.5, -0.5) {};
		\node [style=none] (26) at (-8.5, -0.5) {};
		\node [style=langstate] (27) at (-8.5, 0.75) {\!\!\!{\tt Alice}\!\!\!};
		\node [style=none] (28) at (0, -0.5) {};
		\node [style=none] (29) at (-5.25, -0.5) {};
		\node [style=none] (30) at (-3.25, -0.5) {};
		\node [style=gray dot] (31) at (-5.25, -1.5) {};
		\node [style=none] (32) at (-3.25, -1.5) {};
		\node [style=none] (33) at (0, -1) {};
		\node [style=none] (34) at (-3.25, 0.5) {};
		\node [style=none] (35) at (0, 0.5) {};
		\node [style=langstate] (36) at (-4.25, 0.75) {\tt\!hates\!};
		\node [style=gray dot] (37) at (-3.25, -0.5) {};
		\node [style=none] (38) at (-8.5, 0.5) {};
		\node [style=none] (39) at (-8.5, 0.5) {}; 
		\node [style=none] (40) at (0, 0.5) {};
		\node [style=white dot] (41) at (-8.5, -0.5) {};
		\node [style=none] (42) at (-8.5, -4) {};
		\node [style=white dot] (43) at (0, -0.5) {};
		\node [style=gray dot] (44) at (-5.25, -0.5) {};
		\node [style=none] (45) at (-8.5, 0.5) {};
		\node [style=none] (46) at (0, -0.5) {};
		\node [style=none] (47) at (-5.25, 0.5) {};
		\node [style=none] (48) at (-5.25, -1.5) {};
		\node [style=gray dot] (49) at (-3.25, -1.5) {};
		\node [style=langstate] (50) at (0, 0.75) {\!\!\!{\tt Bob}\!\!\!};
	\end{pgfonlayer}
	\begin{pgfonlayer}{edgelayer}
		\draw [style=swap] (6.center) to (5.center);
		\draw [style=swap] (4.center) to (2.center);
		\draw [style=swap, in=0, out=180, looseness=1.25] (3.center) to (6.center);
		\draw [style=swap] (11.center) to (8);
		\draw [style=swap] (8) to (9);
		\draw [style=swap] (23.center) to (15.center);
		\draw [style=swap] (17.center) to (21.center);
		\draw [style=swap, in=180, out=0, looseness=1.25] (22.center) to (23.center);
		\draw [style=swap] (13.center) to (19);
		\draw [style=swap] (19) to (12);
		\draw [style=swap] (26.center) to (45.center);
		\draw [style=swap] (25.center) to (42.center);
		\draw [style=swap, in=0, out=180, looseness=1.25] (29.center) to (26.center);
		\draw [style=swap] (47.center) to (44);
		\draw [style=swap] (44) to (31);
		\draw [style=swap] (28.center) to (24.center);
		\draw [style=swap] (46.center) to (33.center);
		\draw [style=swap, in=180, out=0, looseness=1.25] (30.center) to (28.center);
		\draw [style=swap] (34.center) to (37);
		\draw [style=swap] (37) to (49);
	\end{pgfonlayer}
\end{tikzpicture}
\eeq
But why do we have to go to quantum circuits to do so?  Indeed, it goes without saying that in absence of quantum circuits we can also compose sentences in order to form larger text, so there should be manner for doing so without having to dive in the realm of the quantum world.

This is indeed the case, when stopping halfway in the middle when passing from language diagrams to quantum circuits \cite{CoeckeText}.  There we  also find circuits that can be composed, and interestingly, some of the features specific to different languages (e.g.~specific orderings of words like subject-verb-object in English) are gone \cite{BVgram}.  So in a manner we end up with something that is more universal than language diagrams, and, for example, stands in a particularly close relationship with visual representations \cite{CoeckeText}.  Here's the language circuit:
\beq
\input{West1.tikz}
\eeq
and here's the corresponding visual representation:
\beq
\input{West3.tikz}
\eeq

Maybe there is something deeper here.  Maybe the origin of language should be looked for in the visual world, rather than in some symbolic abstract realm.  The passage from these visual circuits to the more complex grammatical structure may actually boil down to forcing something that wants to live in two (or more) dimensions on a line, and force additional bureaucracy (or, bureau-crazy?) like the ordering of words, which ends up being different for different languages.  Why forced into one-dimension?  Since it seems to be hard for us humans to communicate verbally in any other way than in terms of one-dimensional strings of sounds.

In fact, the very success of string diagrams is that it allows things to live in two (or more) dimensions. Here's a demonstration of this fact.  Consider the operations of parallel composition $\otimes$ (a.k.a.~`while')  and sequential composition $\circ$ (a.k.a.~`after').   Typically, in textbooks, this structure is represented by one-dimensional strings of algebraic symbols, and then, equations are needed to express their interaction (a.k.a.~bifunctoriality \cite{MacLane}):
\[
(g_1\otimes g_2)\circ(f_1\otimes f_2)=(g_1\circ f_1)\otimes(g_2\circ f_2)
\]
On the other hand, representing it in terms of two-dimensional diagrams we have:
%(see e.g.~\cite{CatsII, CKbook} for details):
\[
\left(\boxmap{g_1}\otimes \boxmap{g_2}\right)\circ\left(\boxmap{f_1}\otimes \boxmap{f_2}\right)
=
\left(\boxmap{g_1}\ \boxmap{g_2}\right)\circ\left(\boxmap{f_1}\ \boxmap{f_2}\right)
=\
\raisebox{0.6mm}{\begin{tikzpicture}
	\begin{pgfonlayer}{nodelayer}
		\node [style=small box] (0) at (0, 0.75) {$f_1$};
		\node [style=small box] (1) at (0, -1) {$g_1$};
		\node [style=none] (2) at (0, 2) {};
		\node [style=none] (3) at (0, -0.75) {};
		\node [style=none] (4) at (0, 0.5) {};
		\node [style=none] (5) at (0, -1.25) {};
		\node [style=none] (6) at (0, -2.25) {};
		\node [style=none] (7) at (0, 1) {};
	\end{pgfonlayer}
	\begin{pgfonlayer}{edgelayer}
		\draw (6.center) to (5.center);
		\draw (7.center) to (2.center);
		\draw (3.center) to (4.center);
	\end{pgfonlayer}
\end{tikzpicture}\  \begin{tikzpicture}
	\begin{pgfonlayer}{nodelayer}
		\node [style=small box] (0) at (0, 0.75) {$f_2$};
		\node [style=small box] (1) at (0, -1) {$g_2$};
		\node [style=none] (2) at (0, 2) {};
		\node [style=none] (3) at (0, -0.75) {};
		\node [style=none] (4) at (0, 0.5) {};
		\node [style=none] (5) at (0, -1.25) {};
		\node [style=none] (6) at (0, -2.25) {};
		\node [style=none] (7) at (0, 1) {};
	\end{pgfonlayer}
	\begin{pgfonlayer}{edgelayer}
		\draw (6.center) to (5.center);
		\draw (7.center) to (2.center);
		\draw (3.center) to (4.center);
	\end{pgfonlayer}
\end{tikzpicture}}
\]
\[
\left(\boxmap{g_1}\circ \boxmap{f_1}\right)\otimes\left(\boxmap{g_2}\circ \boxmap{f_2}\right)
=
\left(\,\raisebox{0.6mm}{\begin{tikzpicture}
	\begin{pgfonlayer}{nodelayer}
		\node [style=small box] (0) at (0, 0.75) {$f_1$};
		\node [style=small box] (1) at (0, -1) {$g_1$};
		\node [style=none] (2) at (0, 2) {};
		\node [style=none] (3) at (0, -0.75) {};
		\node [style=none] (4) at (0, 0.5) {};
		\node [style=none] (5) at (0, -1.25) {};
		\node [style=none] (6) at (0, -2.25) {};
		\node [style=none] (7) at (0, 1) {};
	\end{pgfonlayer}
	\begin{pgfonlayer}{edgelayer}
		\draw (6.center) to (5.center);
		\draw (7.center) to (2.center);
		\draw (3.center) to (4.center);
	\end{pgfonlayer}
\end{tikzpicture}}\right)  \otimes \left(\,\raisebox{0.6mm}{\begin{tikzpicture}
	\begin{pgfonlayer}{nodelayer}
		\node [style=small box] (0) at (0, 0.75) {$f_2$};
		\node [style=small box] (1) at (0, -1) {$g_2$};
		\node [style=none] (2) at (0, 2) {};
		\node [style=none] (3) at (0, -0.75) {};
		\node [style=none] (4) at (0, 0.5) {};
		\node [style=none] (5) at (0, -1.25) {};
		\node [style=none] (6) at (0, -2.25) {};
		\node [style=none] (7) at (0, 1) {};
	\end{pgfonlayer}
	\begin{pgfonlayer}{edgelayer}
		\draw (6.center) to (5.center);
		\draw (7.center) to (2.center);
		\draw (3.center) to (4.center);
	\end{pgfonlayer}
\end{tikzpicture}}\right)
\ =\
\raisebox{0.6mm}{\begin{tikzpicture}
	\begin{pgfonlayer}{nodelayer}
		\node [style=small box] (0) at (0, 0.75) {$f_1$};
		\node [style=small box] (1) at (0, -1) {$g_1$};
		\node [style=none] (2) at (0, 2) {};
		\node [style=none] (3) at (0, -0.75) {};
		\node [style=none] (4) at (0, 0.5) {};
		\node [style=none] (5) at (0, -1.25) {};
		\node [style=none] (6) at (0, -2.25) {};
		\node [style=none] (7) at (0, 1) {};
	\end{pgfonlayer}
	\begin{pgfonlayer}{edgelayer}
		\draw (6.center) to (5.center);
		\draw (7.center) to (2.center);
		\draw (3.center) to (4.center);
	\end{pgfonlayer}
\end{tikzpicture}\  \begin{tikzpicture}
	\begin{pgfonlayer}{nodelayer}
		\node [style=small box] (0) at (0, 0.75) {$f_2$};
		\node [style=small box] (1) at (0, -1) {$g_2$};
		\node [style=none] (2) at (0, 2) {};
		\node [style=none] (3) at (0, -0.75) {};
		\node [style=none] (4) at (0, 0.5) {};
		\node [style=none] (5) at (0, -1.25) {};
		\node [style=none] (6) at (0, -2.25) {};
		\node [style=none] (7) at (0, 1) {};
	\end{pgfonlayer}
	\begin{pgfonlayer}{edgelayer}
		\draw (6.center) to (5.center);
		\draw (7.center) to (2.center);
		\draw (3.center) to (4.center);
	\end{pgfonlayer}
\end{tikzpicture}}
\]
That is, by using the two-dimensional  format of diagrams, the symbolic equation is always satisfied, so it is not needed anymore.  In other words, the equation was a piece of bureaucracy due to the one dimensional symbolic representation.

\section{Meaning-aware relational wholism}\label{sec:wholism}

Let's end with a bit of philosophy, which puts our language diagrams within a broader context of different approaches that one may take towards machine learning.  We'll again do this by means of a metaphor.  Assume one wants to design a home.  One way to do so would be to train a machine using many different images of homes, possibly with some indications of appreciation.  Then, let the machine be creative and design a home based on all of that input.

And it comes up with an amazing design.  But what dit it really do?  It kind of carved everything out of one big block of stuff, and there will be forks and plates on the table, chairs at the table, pictures on the wall, since that's were they were in all of the training images. However, the machine is unaware of what the use is of each of these objects, and in particular that there is a very special relationship between the plate and the fork,  and also with the chair at the table (and less so with the picture on the wall), and even more so, that there is a person who happens to be not in the picture, but who wants to exploit this special relationship with respect to the food they are cooking in the kitchen.

On the other hand, our language diagrams, when applied to a broader context, can capture these relationships, in contrast to the black-box approach that is taken within machine learning:
\beq
\begin{tikzpicture}[scale=0.80]
	\begin{pgfonlayer}{nodelayer}
		\node [style=none] (0) at (-14.75, 6.5) {};
		\node [style=none] (1) at (-14.75, 6.5) {};
		\node [style=langstate] (2) at (-8.75, 6.75) {\!\!\!{\tt table}\!\!\!};
		\node [style=none] (3) at (-8.75, 6.5) {};
		\node [style=none] (4) at (-10.75, 4.75) {};
		\node [style=none] (5) at (-8.75, 6.5) {};
		\node [style=none] (6) at (-12.75, 4.75) {};
		\node [style=langstate] (7) at (-11.75, 5) {\tt\!is at\!};
		\node [style=langstate] (8) at (-14.75, 6.75) {\!\!\!{\tt chair}\!\!\!};
		\node [style=none] (9) at (-14.75, 3) {};
		\node [style=none] (10) at (-14.75, 6.5) {};
		\node [style=white dot] (11) at (-14.75, 3) {};
		\node [style=none] (12) at (-14.75, 3) {};
		\node [style=none] (13) at (-14.75, -7) {};
		\node [style=none] (14) at (-8.75, 6.5) {};
		\node [style=white dot] (15) at (-8.75, 3) {};
		\node [style=none] (16) at (-8.75, 3) {};
		\node [style=none] (17) at (-8.75, 6.5) {};
		\node [style=none] (18) at (-10.75, 4.75) {};
		\node [style=none] (19) at (-8.75, 3) {};
		\node [style=none] (20) at (-8.75, -7) {};
		\node [style=none] (21) at (-8.75, 6.5) {};
		\node [style=none] (22) at (3.25, 6.5) {};
		\node [style=none] (23) at (-2.75, 6.5) {};
		\node [style=none] (24) at (-2.75, 6.5) {};
		\node [style=langstate] (25) at (-2.75, 6.75) {\!\!\!{\tt plate}\!\!\!};
		\node [style=none] (26) at (-2.75, 6.5) {};
		\node [style=none] (27) at (3.25, 6.5) {};
		\node [style=none] (28) at (3.25, 6.5) {};
		\node [style=langstate] (29) at (3.25, 6.75) {\!\!\!{\tt Bob}\!\!\!};
		\node [style=none] (30) at (3.25, 6.5) {};
		\node [style=none] (31) at (3.25, 6.5) {};
		\node [style=none] (32) at (9.25, 6.5) {};
		\node [style=none] (33) at (9.25, 6.5) {};
		\node [style=langstate] (34) at (9.25, 6.75) {\!\!\!{\tt food}\!\!\!};
		\node [style=none] (35) at (9.25, 6.5) {};
		\node [style=none] (36) at (9.25, 6.5) {};
		\node [style=none] (37) at (9.25, 6.5) {};
		\node [style=none] (38) at (3.25, -7) {};
		\node [style=none] (39) at (-2.75, -7) {};
		\node [style=none] (40) at (9.25, -7) {};
		\node [style=white dot] (41) at (-2.75, 1) {};
		\node [style=none] (42) at (-8.75, 1) {};
		\node [style=white dot] (43) at (-8.75, 1) {};
		\node [style=none] (44) at (-4.75, 2.75) {};
		\node [style=none] (45) at (-8.75, 1) {};
		\node [style=langstate] (46) at (-5.75, 3) {\tt\!is on\!};
		\node [style=none] (47) at (-2.75, 1) {};
		\node [style=none] (48) at (-2.75, 1) {};
		\node [style=none] (49) at (-6.75, 2.75) {};
		\node [style=white dot] (50) at (-14.75, -1) {};
		\node [style=white dot] (51) at (3.25, -1) {};
		\node [style=none] (52) at (1.25, 0.75) {};
		\node [style=none] (53) at (3.25, -1) {};
		\node [style=none] (54) at (-0.75, 0.75) {};
		\node [style=langstate] (55) at (0.25, 1) {\tt\!is on\!};
		\node [style=none] (56) at (-0.75, 0.75) {};
		\node [style=none] (57) at (3.25, -1) {};
		\node [style=none] (58) at (-14.75, -1) {};
		\node [style=none] (59) at (-14.75, -1) {};
		\node [style=white dot] (60) at (-2.75, -3) {};
		\node [style=white dot] (61) at (9.25, -3) {};
		\node [style=none] (62) at (7.25, -1.25) {};
		\node [style=none] (63) at (9.25, -3) {};
		\node [style=none] (64) at (5.25, -1.25) {};
		\node [style=langstate] (65) at (6.25, -1) {\tt\!is in\!};
		\node [style=none] (66) at (9.25, -3) {};
		\node [style=none] (67) at (-2.75, -3) {};
		\node [style=none] (68) at (-2.75, -3) {};
		\node [style=white dot] (69) at (9.25, -6) {};
		\node [style=none] (70) at (3.25, -6) {};
		\node [style=none] (71) at (5.25, -4.25) {};
		\node [style=none] (72) at (9.25, -6) {};
		\node [style=langstate] (73) at (6.25, -4) {\tt\!is in\!};
		\node [style=none] (74) at (9.25, -6) {};
		\node [style=none] (75) at (7.25, -4.25) {};
		\node [style=white dot] (76) at (3.25, -6) {};
		\node [style=none] (77) at (3.25, -6) {};
	\end{pgfonlayer}
	\begin{pgfonlayer}{edgelayer}
		\draw [style=swap] (9.center) to (10.center);
		\draw [style=swap] (12.center) to (13.center);
		\draw [style=swap, in=0, out=-90, looseness=0.75] (6.center) to (9.center);
		\draw [style=swap] (19.center) to (14.center);
		\draw [style=swap] (16.center) to (20.center);
		\draw [style=swap, in=180, out=-90, looseness=0.75] (18.center) to (19.center);
		\draw [style=swap] (26.center) to (39.center);
		\draw [style=swap] (31.center) to (38.center);
		\draw [style=swap] (37.center) to (40.center);
		\draw [style=swap, in=0, out=-90, looseness=0.75] (44.center) to (45.center);
		\draw [style=swap, in=180, out=-90, looseness=0.75] (49.center) to (47.center);
		\draw [style=swap, in=0, out=-90, looseness=0.50] (52.center) to (58.center);
		\draw [style=swap, in=180, out=-90, looseness=0.75] (54.center) to (53.center);
		\draw [style=swap, in=0, out=-90, looseness=0.50] (62.center) to (67.center);
		\draw [style=swap, in=180, out=-90, looseness=0.75] (64.center) to (63.center);
		\draw [style=swap, in=0, out=-90, looseness=0.75] (75.center) to (77.center);
		\draw [style=swap, in=180, out=-90, looseness=0.75] (71.center) to (74.center);
	\end{pgfonlayer}
\end{tikzpicture}
\eeq
The kind of thinking encoded in our language diagrams, which is fundamentally wholistic while still aiming to grasp relationships between the parts, is what we mean by the title of this section.

What our language diagrams do is express how words interact with each other.  Once we have established that structure,  one thing that we can do is compute the meaning of a sentence from the meanings of words.  The more interesting thing however, is to derive the meanings of words from the meaning of sentences.  This is exactly what we did in our QNLP experiment: from the knowledge that the sentences in $Train$ were true, we figured out the optimal parameters $\alpha_A, \beta_A,\alpha_B, \beta_B, \alpha, \beta, \ldots$ which determine the meanings of the words.

This is really what this paper was all about, instantiated in order to make qubits speak.  \bM After this philosophy interlude...
%This is what we are aiming for...

\section{...let's get playing!}

First, we need some data to play with. For this we have two options: 1) we
generate an artificial dataset by designing the grammar ourselves, this is the
simplest option and the one we have implemented in \cite{QNLP-foundations} or 2)
we take some real-world text data (e.g. every sentence from Alice in Wonderland)
and use a parser to construct the diagrams automatically for us. In any case,
we begin our experiment with a collection of diagrams for sentences, with some
annotation telling us whether each sentence is true or false (or whether or not
it comes from Alice in Wonderland).

Next, we need to choose an ansatz for each type of word, i.e. an appropriate
circuit shape. Instantaneous quantum polynomials (IQP) \cite{shepherd2009temporally}
are a reasonable choice, since they are shallow circuits but are still believed
to be hard to simulate for classical computers. Once we've chosen the circuit
shape for each type, this defines a \emph{landscape}, the total number of
parameters defining our model: the sum of the parameters defining each word.
For each point in this landscape, i.e. for each combination of parameters, we
can run the circuits corresponding to each sentence in our dataset and check how
far away we are from the correct label. The sum of the distance between our
model's prediction and the correct label is called the \emph{loss function},
this is the objective that we want to minimise.

Once we have defined the landscape and the loss, we
can feed them as input to the optimisation algorithm of our choice. The easiest
choice, and the one we implemented in \cite{QNLP-foundations}, is to pick a
black box optimiser such as simultaneous perturbation stochastic approximation
(SPSA) \cite{spall1992multivariate}. A more advanced solution is to go fully
diagrammatic and compute the gradient of our loss directly from the shape
of the diagrams for sentences \cite{toumi2021diagrammatic}.
When the process has converged to some (locally) optimal parameters, tadam! We
have successfully trained our quantum natural language model. We may now
evaluate it on some testing data, typically a subset of the sentences that we
withhold from the training set. Note that crucially, even though our model has
never seen these sentences during training, hopefully it has already learnt the
meanings for the words, and it knows how to combine these meanings in new ways.

Thus, a typical QNLP experiment may be summarised as follows: 1) we draw the
diagrams for each sentences, 2) we pick a circuit shape for each word type, 3)
we define a loss function, 4) we find the optimal parameters and 5) we test
the model's prediction. These five steps fit in a few lines of Python using
the tools from the DisCoPy library, the result can then be presented in the form
of a Jupyter notebook, see the documentation for plenty of examples:
\begin{center}
\href{https://discopy.readthedocs.io/en/main/notebooks.qnlp.html}{https://discopy.readthedocs.io/en/main/notebooks.qnlp.html}
\end{center}
Also, here is a more detailed tutorial as well:
\begin{center}
\href{https://discopy.readthedocs.io/en/main/notebooks/qnlp-tutorial.html}{https://discopy.readthedocs.io/en/main/notebooks/qnlp-tutorial.html}
\end{center}

\section{Outlook}

So what's next?  Well, as we speak the editor of this volume is himself getting to play, and obviously, we will also embark on a musical adventure.  Evidently, we will also further develop QNLP  bringing in more conceptual depth, for the specific case of language as well as towards broader AI features.  Experimentally, we keep pushing along as hardware improves.

All together this is just the beginning of an adventure that, we hope, on the one hand,  will create the next generation of AI, and on the other hand, will further the quest towards better understanding the human mind and its interaction with the world around us.  \e

\bibliographystyle{plain}
\bibliography{main}
\end{document}